\documentclass[]{elsarticle}
\usepackage{amsmath}
\usepackage{hyperref}
\usepackage{subfig}
\usepackage{graphicx}
\usepackage{epstopdf}
\usepackage{braket}
\usepackage{tikz}

\journal{Journal of \LaTeX\ Templates}
\begin{document}


\begin{frontmatter}

\title{Quantum thermostatted disordered systems and sensitivity under compression}

\author{Tommaso Vanzan$^1$}
\address{$^1$ Section de math\'ematiques, Universit\'e de Gen\`eve, 2-4 rue du Li\`evre,Gen\`eve \\ tommaso.vanzan@unige.ch}

\author{Lamberto Rondoni$^{2,3,4}$}
\address{$^2$ Dipartimento di Scienze Matematiche, Politecnico di Torino, \linebreak Corso Duca degli Abruzzi 24, I-10129 Torino, Italy \\
$^{3}$ \quad INFN, Sezione di Torino, Via P. Giura 1, I-10125 Torino, Italy\\
$^{4}$ \quad Malaysia Italy Centre of Excellence for Mathematical Sciences \\
University Putra Malaysia, {43400~Serdang, Selangor}, Malaysia \\
lamberto.rondoni@polito.it}

\begin{abstract}
A one-dimensional quantum system with off diagonal disorder, consisting of a sample of conducting regions 
randomly interspersed within potential barriers is considered.
Results mainly concerning the large $N$ limit are presented. In particular, the effect of compression on
the transmission coefficient is investigated. A numerical 
method to simulate such a system, for a physically relevant number of barriers, is proposed. It is shown that the disordered model converges to the periodic case as $N$ increases, with a rate of convergence which depends on the disorder degree. Compression always leads to a decrease of the transmission coefficient which may be exploited  to design nano-technological sensors. Effective choices for the physical parameters to improve the sensitivity are provided. Eventually large fluctuations and rate functions are analysed.
\end{abstract}


\end{frontmatter}

\date{\today}

\section{Introduction}
Equilibrium and nonequilibrium thermodynamics \cite{de2013non} are based on the vast separation between the 
space and time scales of the microscopic, mesoscopic and macroscopic physical realms. Such a separation of scales 
requires the systems of interest to be made of very large numbers of microscopic constituents and it allows the state of local thermodynamic equilibrium. In that state, microscopic 
fluctuations of physical quantities are negligible, so that the thermodynamic fields are defined 
and they are described by the thermodynamic laws. In certain {\em small} systems, pertaining {\em e.g.}\ 
to modern bio- and nano-technologies, the separation of scales is not realized, and the 
physical properties of interest are characterized by fluctuations of size comparable to that 
of the average signals. 

In this work, the investigation of Refs.\cite{col1,col2}, concerning a variation of the
Anderson model \cite{anderson,markos2008wave} of disordered solids, is developed in order to account for
the effects of compression on the electron transmission coefficient. Indeed, since Anderson's paper, the study of electron transport has played a key role in the development of modern solid state physic, see for instance Refs. \cite{Celardo1,Celardo2,Celardo3} and references therein.
The systems of interest, here, are nanostructured devices made of an insulating matrix embedding randomly distributed 
drops of conducting material. Such complex objects can be represented by 1-dimensional models 
consisting of conducting regions delimited by $N$ randomly placed potential barriers, in which 
electrons are injected from one electrode at a given temperature $T$ \cite{col1,col2}.
The large $N$ limit is taken under the constraint that the sum of the $N$ barrier widths 
and the total length of the system remain constant as $N$ grows. This is at variance with models that grow in size with $N$.

Unlike usual models found 
in the literature \cite{markos2008wave}, the one of Refs.\cite{col1,col2} enjoys a purely off-diagonal disorder \cite{soukoulis1981off} 
that affects the tunnelling couplings among the wells, but not the energies of the bound states
within the wells. This is not the case of the original tight-binding model introduced by Anderson to describe 
localization phenomena in disordered solids \cite{anderson}, in which random fluctuations 
only concern the energy of a bound state. Furthermore, increasing the number of barriers 
leads, in the Anderson model, to the infinitely large system limit, while increasing 
$N$ in the model investigated here, produces finer and finer distributions of the same amount 
of conductor dispersed within the same amount of insulating material. Therefore, the two large $N$ limits
do not describe the same situation: Anderson's limit views the system of interest as 
macroscopic, i.e. very large compared to its microscopic constituents, while the limit of 
Refs.\cite{col1,col2} refers to system sizes that can be small compared to macroscopic objects. 
The relevant different mathematical constructions imply substantial differences, describing such different physical situations. While Anderson's limit
suits macroscopic objects, the limit of Refs.\cite{col1,col2} better describes systems at 
the mesoscopic scale.

In \cite{col1,col2}, the $N \to \infty$ limit led to the conclusion that a large deviation principle 
holds for the fluctuations of the transmission coefficient, with a proper scaling for the rate function. In the present article, we focus on the behaviour of the transmission coefficient for 
physically relevant numbers of potential barriers, and we study the effects of compression, that can be realized in practice in numerous nanostructured devices.

Our findings are the following:
\begin{itemize}
\item We have extended the continuum limit results proposed in \cite{col1}, observing that the rate of convergence of our model to the Kronig-Penny case \cite{col3} averaged over the energy strongly depends on the disorder degree. 
\item Unlike the Anderson model, our large $N$ limit implies no localization. Nevertheless, increasing the disorder degree at fixed $N$ leads to a substantial reduction of the transmission coefficient, which may be viewed as a phenomenon in some sense analogous to localization.
\item A mathematical framework of compression has been introduced and two different situations have been simulated and compared. In both cases, compression induces a decrease of the transmission coefficient. 
\item Analysing the relative percentage change of the transmission coefficient, an optimal configuration has been identified to design an effective sensor. It is found that a moderate number of barriers and strong disorder imply high sensitivity to compression.
\item Fluctuations and rate function have been investigated, obtaining that they may be exploited to reveal the compression state of the system.
\item A numerical scheme which does not suffer from overflow and $\Omega$ problem has been developed.
\end{itemize}
This article is organized as follows: Section 2 describes the mathematical model to be used for disorder 
and compression. Section 3 introduces our numerical results and it is divided in subsections concerning linear compression model and a generalized version. Section 4 deals with fluctuations 
and rate functions for systems under compression. Section 5 recapitulates the contents of the article and in the appendix the numerical scheme developed to tackle the issues raised by the range of energies and lengths of physical interest is explained.
\section{The model}
\begin{figure}
    \begin{tikzpicture}[xscale=3.6,yscale=0.8]
      \draw [->] (-.1,1) -- (3.4,1) node[right, scale = 1.3] {$x$};
      \draw [->] (0,1) node[below]{$x_0$} -- (0,4);
      \node at (-0.15,3) {\small{$V$}};
      \draw [thick] (0.3,1) node[below]{$x_1$} -- (0.3,3);
      \draw [thick] (0,3) node[below]{} -- (0.3,3);
      \draw [thick] (0.7,1) node[below]{$x_{2}$} -- (0.7,3);
      \draw [thick] (1.2,1) node[below]{$x_{3}$} -- (1.2,3);
      \draw [thick] (0.7,3) node[below]{} -- (1.2,3);
	 \draw [thick] (1.9,1) node[below]{$x_{2N-4}$} -- (1.9,3);      
      \draw [thick] (2.5,1) node[below]{$x_{2N-3}$} -- (2.5,3);
      \draw [thick] (1.9,3) node[below]{} -- (2.5,3);
      \draw [thick] (2.9,1) node[below]{$x_{2N-2}$} -- (2.9,3);      
      \draw [thick] (3.2,1) node[below]{$x_{2N-1}$} -- (3.2,3);
		\draw [thick] (2.9,3) node[below]{} -- (3.2,3);
	 \node at (1.55,1.4) {\small{$\cdots$}};
      \node at (1.55,1.7) {\small{$\cdots$}};
     \node at (0.15,1.4) {\small{$\lambda_1$}};
     \node at (0.95,1.4) {\small{$\lambda_2$}};
     \node at (2.2,1.4) {\small{$\lambda_{N-1}$}};
     \node at (3.05,1.4) {\small{$\lambda_{N}$}}; 
     \node at (0.5,1.4) {\small{$\delta_N$}};
      \node at (2.7,1.4) {\small{$\delta_N$}};
 \end{tikzpicture}
    \caption{1D multiple-well system, consisting of: {\bf a)} $N$ potential barriers, whose width $\lambda_j$ is uniformly randomly distributed; {\bf b)} $(N - 1)$ conducting regions of width $\delta_N$; {\bf c)} left boundary representing a classical thermostat at temperature $T$; {\bf d)} empty space as right boundary.}
    \label{sistema}
\end{figure}
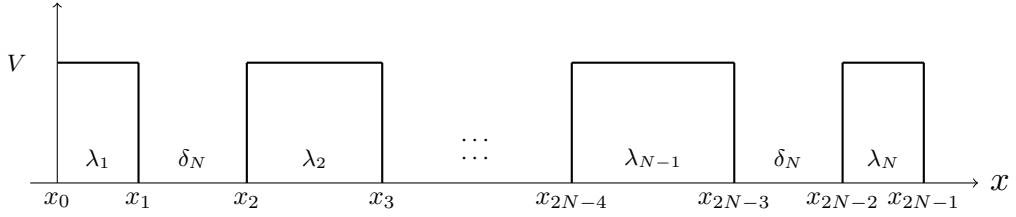

Consider a 1-dimensional system of length $L$, consisting of an array of $N$ potential barriers 
separating $N-1$ potential wells, in equilibrium with one electrode that acts as an external thermostat 
at temperature $T$, cf.\ Fig.\ref{sistema}. This means that the mean energy of the plane waves entering 
from the left boundary is $k_BT/2$.
Let the wells have same width $\delta_N$, so that the total 
length of the $N-1$ wells is $L_{cond} = (N-1) \delta_N = \alpha L$, where $\alpha \in (0,1)$, and let 
the widths of the $N$ potential barriers be picked at random with uniform distribution, to reach the 
total length $(1-\alpha) L$ (cf.\ section \ref{sec:sectiondisorder} for details).

Let all potential barriers have same constant height $V (x) = V$, and let their boundary points be 
denoted by $x=x_0,...,x_{2N-1}$. For fixed barrier width, we would have a variation of the Kronig-Penney model 
\cite{col3}. 
In a steady state, the microscopic behavior of the electrons in this environment is given by the 
time independent Schr\"{o}dinger equation:
\begin{equation}\label{Schrodinger}
\frac{d^2}{dx^2}\psi=\frac{2m}{\hbar^2}(V-E)\psi,\hspace{1cm}x\in[0,L]
\end{equation}
where $m$ is the mass of an electron, and $\hbar$ is the reduced Planck constant.
Denoting by $U_l$ the $l$-th region, for $l\in \left \lbrace 0,2,...,2N \right \rbrace$,  
the solutions of eq.\eqref{Schrodinger} for $E<V$ have the form:
\begin{equation}\label{solution}
\psi_l(x)=
\begin{cases}
A_{2l}e^{ikx}+A_{2l+1}e^{-ikx}\hspace{0.5cm}  \mbox{for $x\in U_l$ and even $l$ 
\hspace{0.1cm}({\em i.e.}\ for}\hspace{0.1cm} V(x)=0)\\
A_{2l}e^{-zx}+A_{2l+1}e^{zx}\hspace{0.73cm} \mbox{for $x\in U_l$ and odd $l$ 
\hspace{0.1cm}({\em i.e.}\ for}\hspace{0.1cm} V(x)=V)
\end{cases}
\end{equation}
with $k=\sqrt{2mE}/\hbar$ and $z=\sqrt{2m(V-E)}/\hbar$.
The boundary conditions prescribe $A_0>0$ for the amplitude of the plane wave entering from the 
left boundary, and $A_{4N+1}=0$ since no wave enters or is reflected from the right boundary.
The steady state current is defined by \cite{schwabl},
\begin{equation}\label{current}
j_l(x)=\frac{\hbar}{2mi}\left[\psi_l(x)^*\left(\frac{d}{dx}\psi_l(x)\right)-
\left(\frac{d}{dx}\psi_l(x)^*\right)\psi_l(x)\right]
=j^{tr}_l(A_{2l})-j^{ref}_l(A_{2l+1}),
\end{equation}
where $j^{tr}_{l}(A_{2l})=\hbar k|A_{2l}|^2 /m$ denotes the current transmitted from the $(l-1)$-th barrier 
on the left and $j^{ref}_l(A_{2l+1})=\hbar k|A_{2l+1}|^2/m $ denotes the current reflected from the $(l+1)$-th 
barrier. 
Considering eqs.\eqref{solution} and \eqref{current}, we get the following definition 
for the transmission coefficient $S$ across the system:
\begin{equation}\label{transmission}
S(N)=\frac{j^{tr}_{2N}(A_{4N})}{j^{tr}_0(A_0)}=\frac{|A_{4N}|^2}{|A_{0}|^2}.
\end{equation}

To numerically compute the coefficient $S$ as a function of the various parameters of the model, 
it is convenient to rewrite eq.\eqref{Schrodinger} in terms of the characteristic quantities, 
introducing $\hat{x}=x/L$, $\hat{\psi}=\psi \sqrt{L}$, $\hat{E}=E/E_T$ and $\hat{V}=V/E_T$, 
with $E_T=K_bT$, which is twice the mean kinetic energy of the plane waves entering from the left 
thermostat. Further, introducing the scalar parameter $\gamma=\hbar^2 /(2mL^2E_T)$, the expression 
for the dimensionless wave vectors takes the form: $\hat{k}={\sqrt{\hat{E}}}/{\sqrt{\gamma}}$ 
and $\hat{z}=\sqrt{\hat{V}-\hat{E}}/{\sqrt{\gamma}}$. 

In the following, we refer only to dimensionless quantities, but for sake of simplicity, we omit the 
hat over the corresponding symbols. Hence, the dimensionless form of eq.\eqref{Schrodinger} reads:
\begin{equation}\label{Schrodingeradime}
\frac{d^2}{dx^2}\psi(x)=\frac{1}{
\gamma}(V-E)\psi(x),\hspace{1.5cm}x\in [0,1].
\end{equation}

\subsection{\label{sec:sectiondisorder}Mathematical treatment of disorder}
We introduce disorder in our systems by picking the dimensionless potential barrier widths, $\hat{\lambda}_i$, 
$i=1,...,N$, from a given probability distribution $\rho(\lambda)d\lambda$. We begin with a uniform distribution: 
$$
\rho(\lambda) = {1 \over 1 - 2 \eta} ~, \quad \lambda \in [\eta,1-\eta] ~, \quad \eta \in (0, 1/2),
$$
where, for a given $L$, $\eta$ is chosen in order to avoid physical nuisances, such as barriers widths smaller 
than single atoms. The smaller is $
\eta$, the larger is the support of the probability density function $\rho(\lambda)$, thus a measure of the disorder degree is given by the value of $\eta$.
The empirical mean width for a single realization of the disorder is a random variable denoted by:
\begin{equation}\label{mean}
\hat{\lambda}_{N} = \frac{1}{N}\sum_{i=1}^{N} \hat{\lambda}_i.
\end{equation}
The weak law of large numbers implies that $\hat{\lambda}_{N}$ converges in probability to the mean 
$\langle \hat\lambda\rangle$, in the large $N$ limit.

After the $N$ widths have been generated, the total length of the sample may exceed or be
smaller than the desired value, therefore we rescale all lengths introducing the parameter 
\begin{equation}
c_N=\frac{L(1-\alpha)}{N\hat{\lambda}_N}
\end{equation}
so that $\lambda_i=c_N\hat{\lambda}_i$ and
\begin{equation}\label{rescale}
\sum_{i=1}^N\lambda_i=\sum_{i=1}^N\frac{L(1-\alpha)}{N\hat{\lambda}_N}\hat{\lambda}_i=L(1-\alpha)=L_{ins}
\end{equation}
Let us denote by $\Lambda_N=\left\lbrace \lambda_1,...,\lambda_N\right\rbrace$ the set of barrier widths. Among the possible realizations of $\Lambda_N$, the regular barrier distribution 
$\Lambda_B=\left\lbrace \lambda_B,...,\lambda_B\right \rbrace$ plays a crucial role, since it corresponds 
to the Kronig-Penney model, the continuum limit of which has been considered in Ref.\cite{col3}. We call
\textit{periodic} the case of $\Lambda_B$.

Considering an observable $A$, defined as a function of a given realization of barriers, and denoting by $\Omega=\left\lbrace \Lambda_N^{(1)},\Lambda_N^{(2)},....,\Lambda_N^{(\ell)}\right\rbrace$ a set of 
realizations, the corresponding ensemble average is given by:
\begin{equation}
\langle A\rangle_{\Omega}=\frac{1}{\ell}\sum_{i=1}^\ell A\left(\Lambda_N^{(i)}\right).
\end{equation}
We are interested in the observable $S$, which is also a function of the energy $E$ of the incoming particle,
of the potential height $V$ and of the temperature $T$ that determines the distribution of the particles
energies: $S=S(\Lambda_N;V,E,T)$. 
Averaging over the particles energy gives the coefficient
\begin{equation}
S(\Lambda_N;V,T)=\int_{0}^{\infty}S(\Lambda_N;E,V,T) f_{eq}(E)dE,
\end{equation}
where the Maxwellian probability density
\begin{equation}
f_{eq}(E)=\sqrt{\frac{1}{\pi E}}e^{-E}
\end{equation}
is used to represent the electrode on the left as a classical heat reservoir.

\subsection{\label{sec:compressionesection}Sample compression}
Because of externally exerted pressure, the sample length may be reduced by an amount $\Gamma$, so that
its length is given by $L_{compr} = L - \Gamma$. If the insulator is {\em e.g.}\ polymeric and the conductor 
is {\em e.g.}\ metallic,
we may in first approximation assume that the length reduction only concerns the potential barrier widths.
In any event, introducing the ratio ${r}$ for the effect of compression 
on the two materials, we may write:
\begin{eqnarray}\label{equazionicompressione}
L_{ins,compr}=L_{ins}-\Gamma\cdot r\\
L_{cond,compr}=L_{cond}-\Gamma\cdot (1-r)\\\
L_{ins,compr}+L_{cond,compr}=L_{compr}
\end{eqnarray}
where the index $compr$ denotes the lengths regarding the compressed state. For instance, the case $r=1$ describes the situation in which only the insulator is affected by the compression.
Introducing the parameter $\alpha_{compr}={L_{cond,compr}}/{L_{compr}}$, the compressed state can 
be described by the function 
\begin{equation}
f_{\Gamma,r}: (L,\alpha,\Lambda_{N},V)\rightarrow (L_{compr},\alpha_{compr},\Lambda_{N,compr},V_{compr})
\end{equation}
that associates the old system, characterized by $(L,\alpha,\Lambda_{N},V)$ with the compressed system 
characterized by $(L_{compr},\alpha_{compr},\Lambda_{N,compr},V_{compr})$, where the notation indicates
that the compression modifies the realization of the barrier widths and, consequently, that it may 
affect the potential height. 

One possibility for the variation of the potential under compression is that the area under a barrier, 
{\em i.e.}\ barrier width times barrier height, is constant. 
The idea is that the compression leads to higher insulator density, hence to an increase of the potential. 
The specific form of the increase is irrelevant here, since other rules may be simply implemented in our
framework.

One may ask whether the compression introduces disorder also in the potential strength, because of
different increments in barriers of different widths. Using our rule, this does not happen.
Indeed, consider a system composed by two barriers of width $\lambda_1$ and $\lambda_2$ 
divided by a conduction region whose length is $\delta$. One has
\begin{equation}
L=\lambda_1+\lambda_2+\delta ~,\qquad L_{ins}=\lambda_1+\lambda_2 ~, \qquad L_{cond}=\delta ~, 
\qquad \alpha=\frac{\delta}{\lambda_1+\lambda_2+\delta}
\end{equation}
Compressing the system by a quantity $d$, and distributing the compression with ratio $r$, one gets:
\begin{eqnarray}
&& L_{compr}=\lambda_1+\lambda_2+\delta-d ~, \quad L_{ins,compr}=\lambda_1+\delta-d\cdot r ~, \\
&& L_{cond,compr}=\delta-d\cdot(1-r) ~, \quad \alpha_{compr}=\frac{\delta-d\cdot(1-r)}{\lambda_1+\lambda_2+\delta-d}
\end{eqnarray}
Observe that the widths $\lambda_1$ and $\lambda_2$ arise from the normalization of realizations 
$\hat \lambda_1$ and $\hat \lambda_2$
picked at random from the chosen distribution of widths. Then, we may write
\begin{equation}\label{compressions}
\lambda_i = \frac{L(1-\alpha)}{N\lambda_N}\hat{\lambda}_i  ~, \quad i=1,2
\end{equation}
If the area under each barrier is kept constant under compression, we have
\begin{equation}
\lambda_i V = \lambda_{i,compr} V_{i,compr} ~, \quad i=1,2
\end{equation}
which, thanks to eq.\eqref{compressions} can be rewritten as:
\begin{equation}
\frac{L(1-\alpha)}{N\lambda_N}\hat{\lambda}_i V = \frac{L_{compr}(1-\alpha_{compr})}{N\lambda_N}\hat{\lambda}_i V_{i, compr}
 ~, \quad i=1,2
\end{equation}
This implies:
\begin{equation}\label{formulacompressione}
V_{1,compr}=V_{2,compr}=V \frac{L(1-\alpha)}{L_{compr}(1-\alpha_{compr})}
\end{equation}
The reasoning can be easily extended to any numbers of barriers.
It follows that the heights of the potential barriers depend only on the compression level
and on the ratio $r$, not on the realization of the microscopic disorder.

\section{Numerical results} \label{sezioneprimanumerica}
The solution \eqref{solution} of eq.\eqref{Schrodinger} must be subjected to the classical BenDaniel-Duke 
boundary conditions on the generic $l$-th node, with $l\in \left \{0,1,...,2N-1\right \}$,
which require the continuity both of the wave function and of its first derivative at each node:
 \begin{equation}\label{condizioni}
$$$$ 
\begin{cases}
\psi_l(x_l)  =\psi_{l+1}(x_l)\\
\psi^{'}_{l}(x_l)  =\psi^{'}_{l+1}(x_l)
\end{cases}
\end{equation}
where $x_l= \sum_{i=1}^{(l/2)} \lambda_i+\delta \frac{l}{2}$, if $l$ is even, and $x_l=\sum_{i=1}^{(l+1)/2} \lambda_i + \delta \frac{l-1}{2}$ if $l$ is odd, where $\lambda_i$ denotes the random width of the i-th barrier. With this notation, eq.\eqref{condizioni} may be written as:
$$
\mathbf{M}_0(x_0) \cdot \begin{pmatrix}
A_0\\A_1\end{pmatrix} = \mathbf{M}_1(x_0) \begin{pmatrix}
A_2\\A_3\end{pmatrix}$$
$$  \mathbf{M}_2(x_1) \cdot \begin{pmatrix}
A_2\\A_3\end{pmatrix} = \mathbf{M}_3(x_1) \begin{pmatrix}
A_4\\A_5\end{pmatrix}$$
$$\mathbf{M}_4(x_0) \cdot \begin{pmatrix}
A_4\\A_5\end{pmatrix} = \mathbf{M}_5(x_0) \begin{pmatrix}
A_6\\A_7\end{pmatrix}
$$
where the support matrices $\mathbf{M}_{2l}$ and $\mathbf{M}_{2l+1}$ have been introduced, and
\begin{equation}\mathbf{M}_{4N-2}(x_{2N-1}) \cdot \begin{pmatrix}
A_{4N-2}\\A_{4N-1}\end{pmatrix} = \mathbf{M}_{4N-1}(x_{2N-1}) \begin{pmatrix}
A_{4N}\\A_{4N+1}\end{pmatrix}
\end{equation}
For $E<V$, these 2x2 matrices of coefficients $\mathbf{M}_{2l}(x_l)$ and $\mathbf{M}_{2l+1}(x_l)$ read:
\begin{equation}
\mathbf{M}_{2l}(x_l)=\begin{pmatrix}
e^{ikx_l} & e^{-ikl}\\ ike^{ikx_l} & -ike^{-ikl}
\end{pmatrix}\hspace{0.2cm} \mbox{and} \hspace{0.2cm}\mathbf{M}_{2l+1}(x_l)=\begin{pmatrix}
e^{-zx_l} & e^{zx_l}\\ -ze^{-zx_l} & ze^{zx_l}
\end{pmatrix}
\end{equation}
for even $l$, and 
\begin{equation}
\mathbf{M}_{2l}(x_l)=\begin{pmatrix}
e^{-zx_l} & e^{zx_l}\\ -ze^{-zx_l} & ze^{zx_l}
\end{pmatrix}\hspace{0.2cm} \mbox{and} \hspace{0.2cm}\mathbf{M}_{2l+1}(x_l)=\begin{pmatrix}
e^{ikx_l} & e^{-ikx_l}\\ ike^{ikx_l} & -ike^{-ikx_l}
\end{pmatrix}
\end{equation}
for odd $l$. Assuming that the amplitude of the incoming wave $A_0$ is known, and imposing $A_{4N+1}=0$, 
since there is no reflection at the right boundary, these equations constitute a set of $4N$ equations 
in $4N$ variables, for which the support matrices $\mathbf{M}_{2l}$ and $\mathbf{M}_{2l+1}$ allow us to write:
\begin{equation}\label{transfermatrices}
\begin{pmatrix}
A_0\\ A_1
\end{pmatrix}=\mathbf{M}_{0}^{-1}\cdot\mathbf{M}_{1}\cdot\mathbf{M}_{2}^{-1}\cdot\mathbf{M}_{3}\cdot\cdot\cdot\mathbf{M}_{4N-2}^{-1}\mathbf{M}_{4N-1}\cdot\begin{pmatrix}
A_{4N}\\0
\end{pmatrix}=\mathbf{M}\begin{pmatrix}
A_{4N}\\0
\end{pmatrix}
\end{equation}
where $\mathbf{M}$ denotes the product of the $\mathbf{M}_i$. It follows that 
\begin{equation}
A_0={M}_{11}A_{4N}
\end{equation}
where $M_{11}$ is the first entry of {\bf M}. Consequently, eq.\eqref{transmission} may be written as:
\begin{equation}\label{transfermatrixtransmission}
S=\frac{A^*_{4N}A_{4N}}{A^*_0A_0}=\frac{1}{|{M}_{11}|^2}
\end{equation}
which is, in principle, a simple and efficient expression for the transmission coefficient. 
In practice, however, the range of energies and lengths of nanotechnological interest make 
eq.\eqref{transfermatrixtransmission} hardly of any use for numerical calculations. 
For instance, $L=500nm$ and energy of the order of $E_T$ at room temperature imply that the dimensional variable $z$ ranges between 
100 and 1000, which make overflow the entries of the matrices $\mathbf{M}_{i}$, see {e.g.}\ Ref.\cite{perez}
for overflow and $\Omega$ problems. To overcome these difficulties, we have developed a numerical scheme which relies uniquely upon the scattering matrix, and that is described
in the Appendix.
\begin{figure}[h]
\centering
\subfloat[][\emph{}]
{{\includegraphics[width=.485\textwidth]{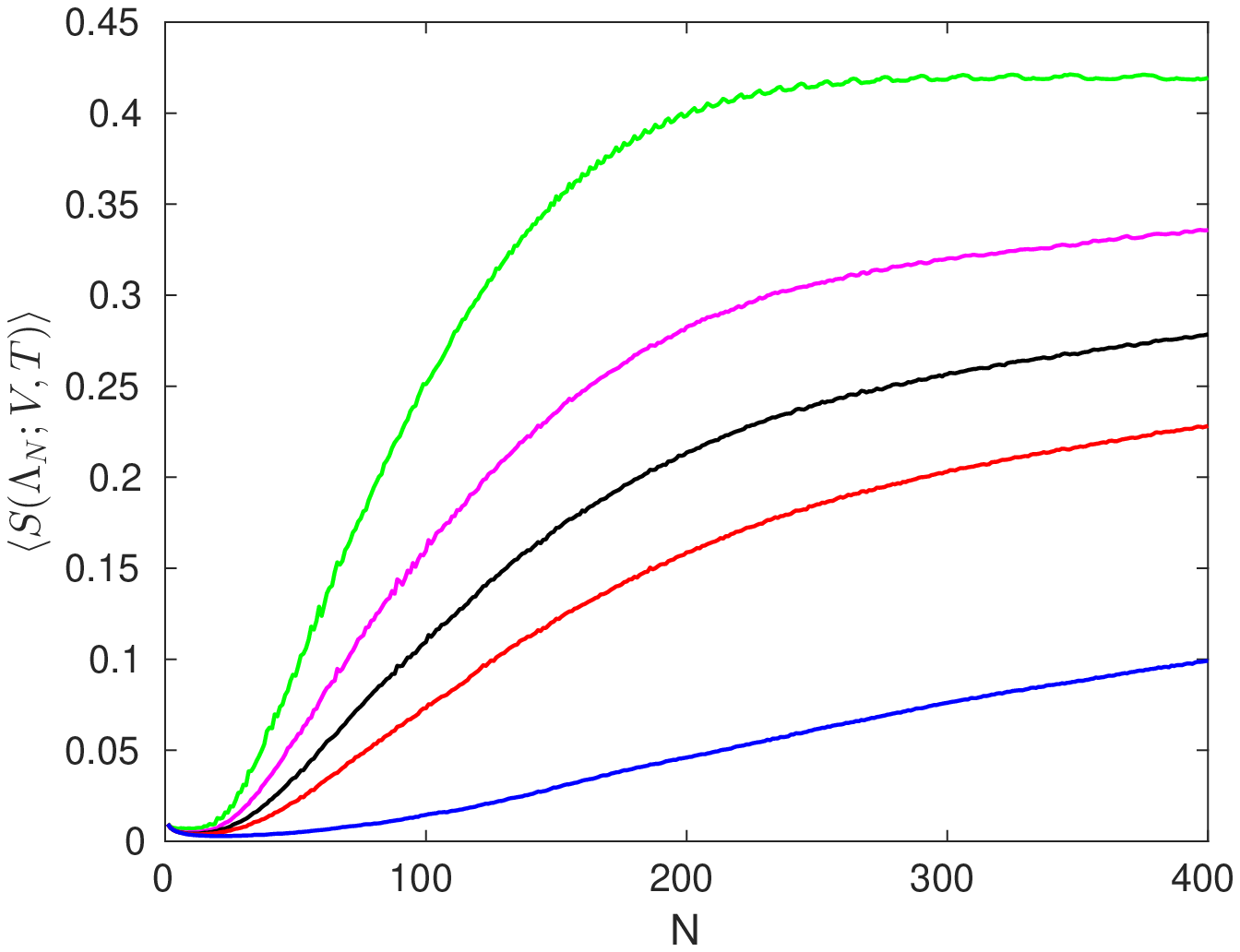}}} \quad
\subfloat[][\emph{}]
{{\includegraphics[width=.485\textwidth]{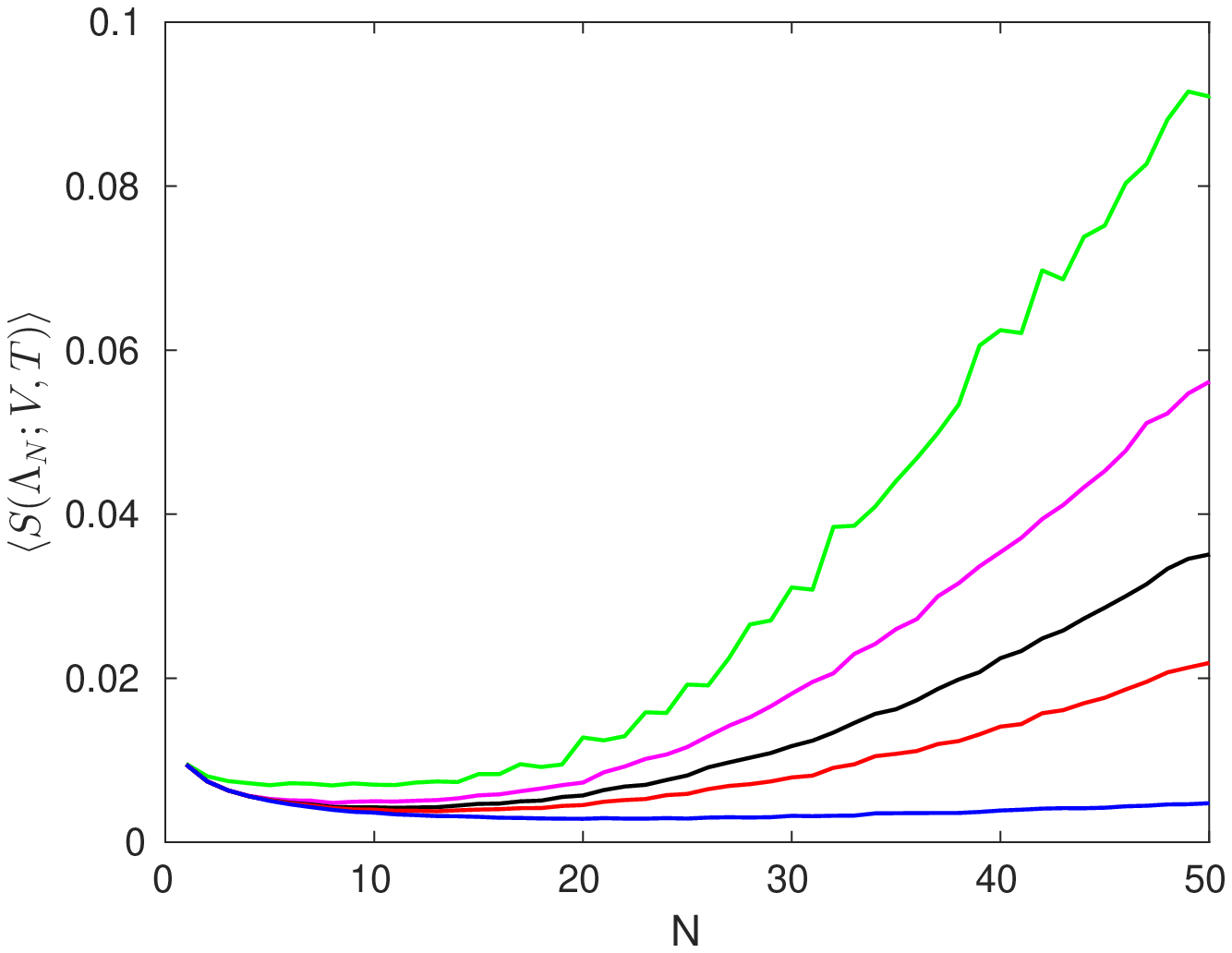}}}\quad
\subfloat[][\emph{}]
{{\includegraphics[width=.485\textwidth]{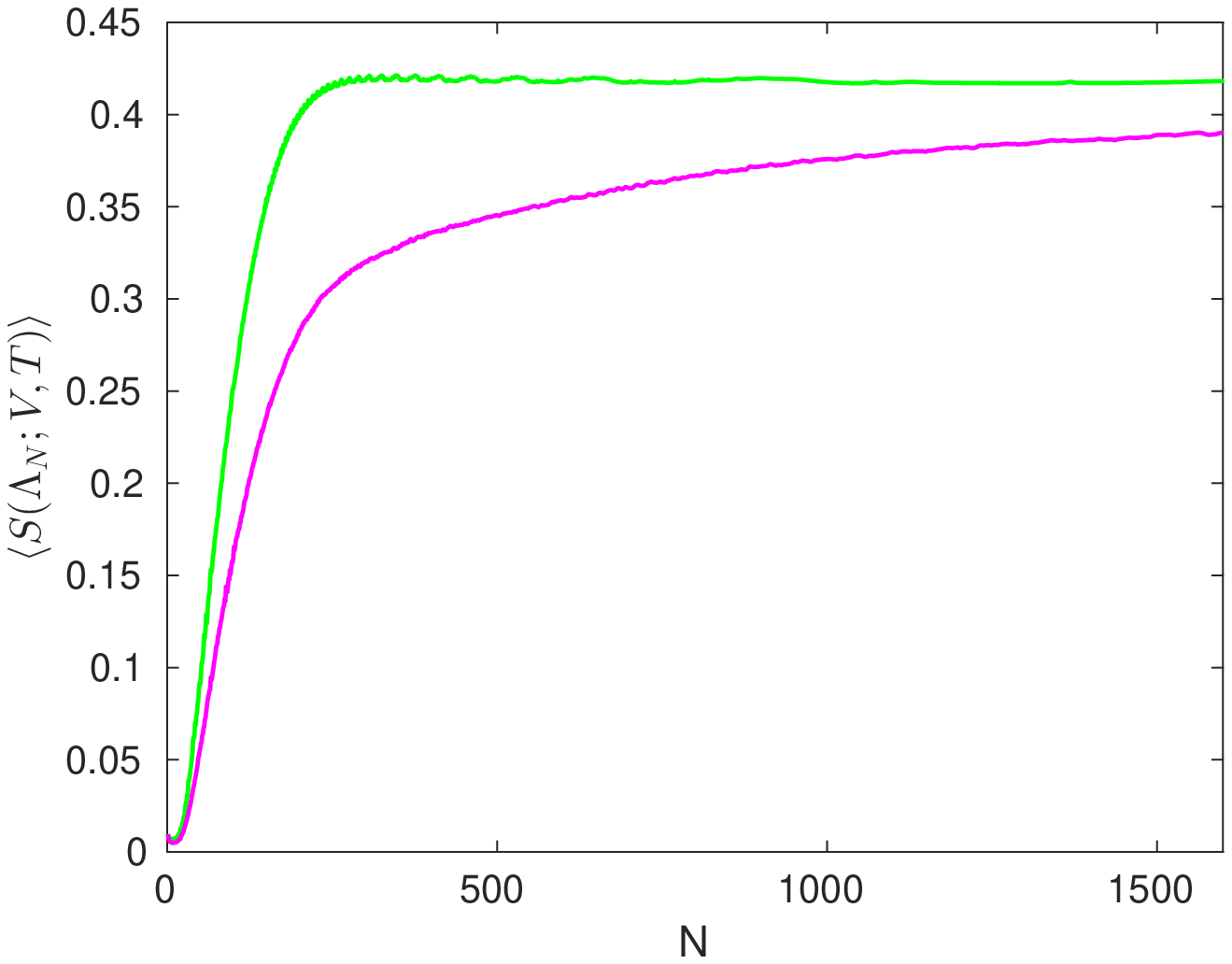}}}
\caption{Left panel: Behavior of $\langle S(\Lambda_N; V, T) \rangle_{\Omega}$ for different barrier distributions. The green line refers to the periodic case, the magenta line to uniform $\rho(\lambda)$ in [0.4,0.6], the black line to uniform $\rho(\lambda)$ in [0.35,0.65], the red line to  uniform $\rho(\lambda)$ in [0.3,0.7], the blu line to uniform $\rho(\lambda)$ in [0.1,0.9]. Right panel: Zoom for small value of $N$ of the left panel figure. Low panel: comparison of the asymptotic behaviour between the periodic case and the random one with $\rho(\lambda)$ in [0.4,0.6].}
\label{comportamentoprobabilita}
\end{figure}
For our numerical results, if not otherwise stated, we refer to $L=500nm$, which is a length suitable for 
present nanotechnology, to $V=3$ for the dimensionless potential, and to $\alpha=10/11$, meaning that the 
insulator length amounts to the fraction $1/11$ of the total sample length.

Figure \ref{comportamentoprobabilita} shows the common behaviour of the ensemble average 
$\langle S(\Lambda_N; V, T) \rangle_\Omega$ as a function of the number of barriers $N$, computed over 
different realizations of the microscopic disorder. The maximum value $N=400$ is determined by the fact that for 
$L=500nm$, one obtains barrier widths of the order of $10^{-10}m$, below which the physical significance is lost. The right panel of Figure \ref{comportamentoprobabilita} concerns the behaviour of $\langle S(\Lambda_N;V,T)\rangle_{\Omega}$ for small values of $N$.

Let us understand as greater disorder the situation in which the support of the uniform distribution of widhs $\rho$ 
is wider, {\em i.e.}\ the case in which $\eta$ is smaller. Then, Fig.\ref{comportamentoprobabilita} shows that the 
periodic case enjoys the highest transmission coefficient, and that growing disorder implies a 
decay of $\langle S \rangle_\Omega$.

At the same time, the growth of $N$ at fixed disorder degree makes $\langle S \rangle_\Omega$ increase, apart
from a minimal decrease at small $N$. The periodic case, in particular, reaches a plateau at $N\approx 200$;
in other words the periodic case attains within physically relevant scales the maximum transmission 
coefficient that the model allows and that remains throughout the physically relevant range.
The disordered cases, on  the other hand, may also reach a plateau, but presumably at scales that 
exceed the physically relevant ones. Therefore, in their cases, larger $N$, {\em i.e.}\ finer
structures, correspond to higher $\langle S \rangle_\Omega$.

This statement agrees with Ref.\cite{col3}, in which a closed formula for the asymptotic behavior of the transmission coefficient in the periodic case has been given:
\begin{equation}
\hat{S}=\lim_{N\rightarrow \infty} S_N= \left[
1+\frac{\tilde{E}^2}{4E}\left(\frac{\sin(L\sqrt{E-\tilde{E}})}{\sqrt{E-\tilde{E}}} \right)^2 \right]^{-1}
\end{equation}
As we are interested in the average with respect to the energy distribution, we numerically computed
\begin{equation}\label{averagemedioE}
S_B = \int_{E}  f_{eq}(E) \left[
1+\frac{\tilde{E}^2}{4E}\left(\frac{\sin(L\sqrt{E-\tilde{E}})}{\sqrt{E-\tilde{E}}} \right)^2 \right]^{-1}
\end{equation}
for different disorder intensities. Setting the parameters given at the beginning of this section we get 
$S_B=0.4178$, while $\langle S(\Lambda_{4300};V,T)\rangle_\Omega = 0.4076$, and 
$\langle S(\Lambda_{10^5};,V,T)\rangle_\Omega = 0.4150$, with $\lambda \in [0.4,0.6]$. 
For $\lambda \in [0.1,0.9]$, we get instead $\langle S(\Lambda_{7\cdot 10^4};V,T)\rangle_\Omega = 0.4073$  

We conclude that in the large $N$ limit our model tends to the Kronig Penney model, with a rate of convergence that 
depends on the disorder degree. This confirms the results of Ref.\cite{col1},
although for highly disordered cases the asymptotic properties do not suit the nanotechnological interests. 
These observations mean that there are no localization effects in our model, unlike the case of 
the Anderson model. The origins of this discrepancy may be traced back to the fact that Anderson's
model is based on a discrete tight binding Hamiltonian, that we do not have, and to the inapplicability 
in our model of Furstenberg's theorem, from which localization depending on the 
first Lyapunov exponent follows \cite{vulpiani},\cite{izrailev2012anomalous}. While the sequence of barriers of Anderson's model 
increases by adding new barriers without modifying the previous ones, adding a barrier
in our construction alters the preceding barriers in order to keep unchanged the insulator amount, cf.\
eq.\eqref{rescale}. The hypothesis of Furstenberg's theorem are thus violated and we are in a framework
that has been little investigated so far.
\subsection{\label{sec:linearcompression}Linear compression model and design optimization for sensor devices}
Suppose now that our samples have been compressed according to the model described in section \ref{sec:compressionesection}. 
\begin{figure}[h]
\centering
\subfloat[][\emph{}]
{\includegraphics[width=.47 \textwidth]{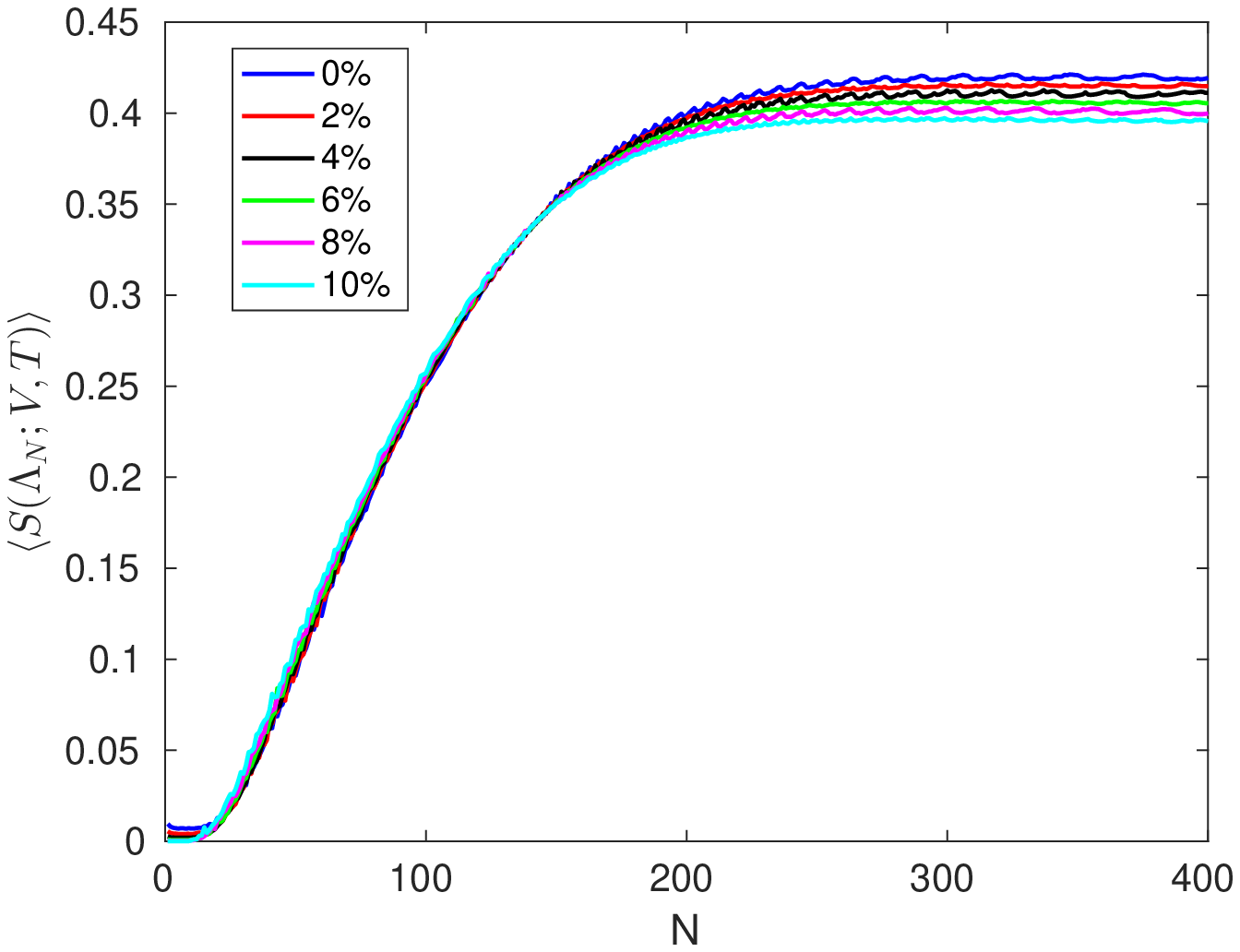}} \quad
\subfloat[][\emph{}]
{\includegraphics[width=.47  \textwidth]{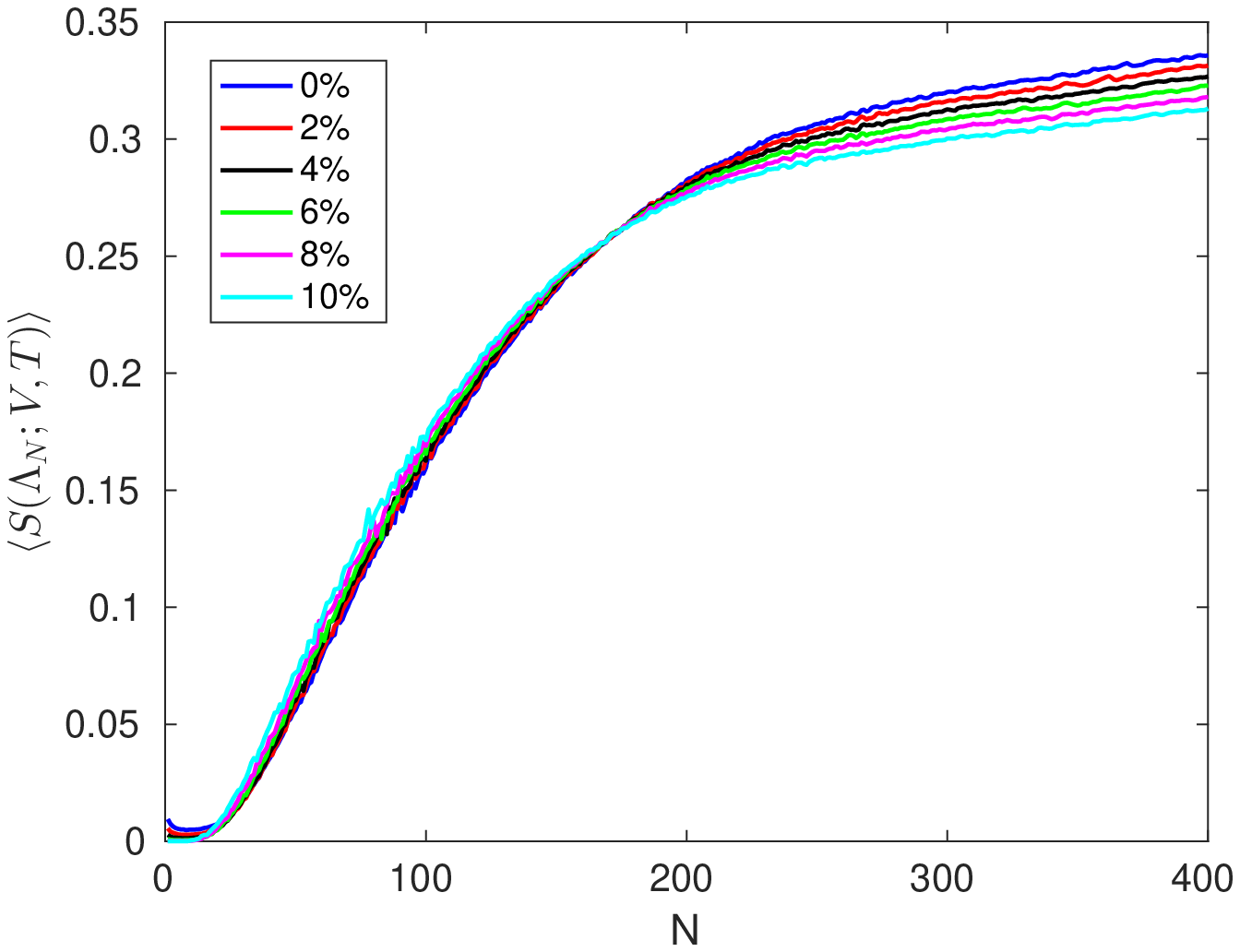}} \quad
\subfloat[][\emph{}]
{\includegraphics[width=.47  \textwidth]{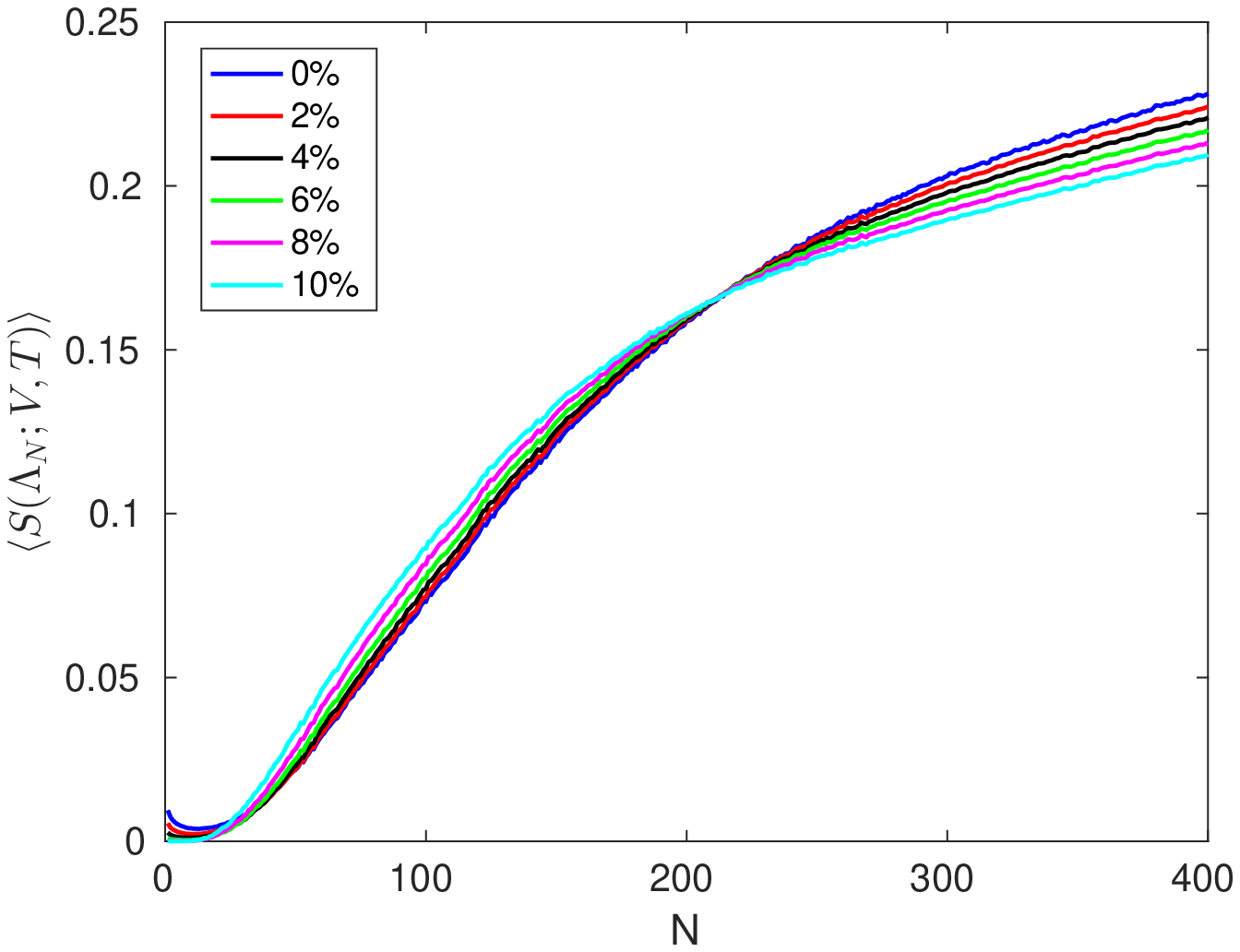}} \quad
\subfloat[][\emph{}]
{\includegraphics[width=.47 \textwidth]{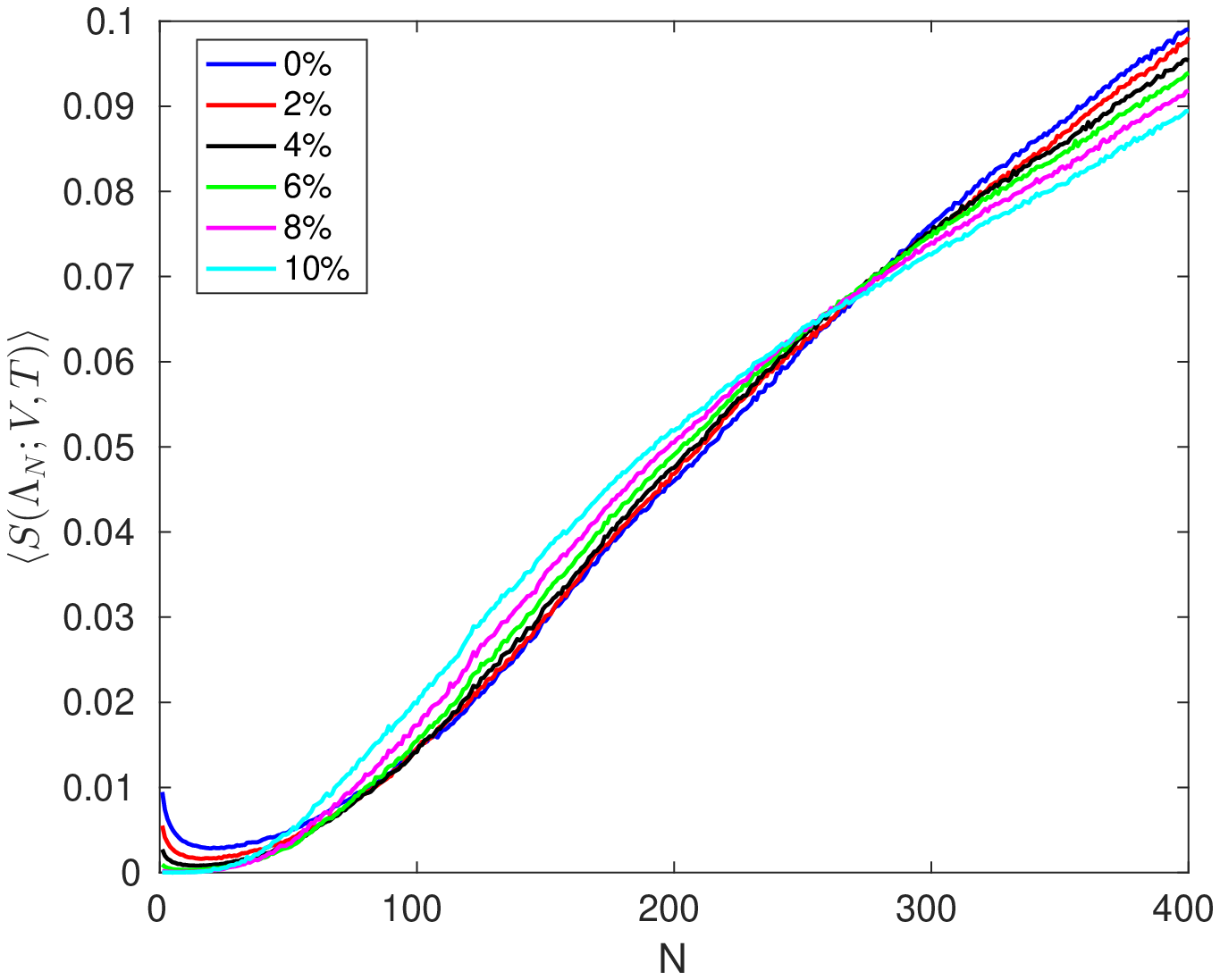}} \quad
\caption{Behavior of $\langle S(\Lambda_N;V,T) \rangle$ subjected to the first model of compression, with $r=0.6$, for a serie of compression percentages.
a) Periodic case b) $\lambda\in[0.4,0.6]$ c) $\lambda\in[0.3,0.7]$ d) $\lambda\in[0.1,0.9]$}
\label{compressionifigure}
\end{figure}

Figure \ref{compressionifigure}
shows that increasing the compression percentage leads in our model to a mild decrease of $S$, 
for large $N$, and to an equally mild increase for small $N$. The cross-over between the two regimes grows with the disorder. Note that the growth of the disorder also seems to move forward,
away form the physically interesting region, the asymptotic regime. 
 \begin{figure}[ht]
\subfloat[][\emph{}]
{\includegraphics[width=.46  \textwidth]{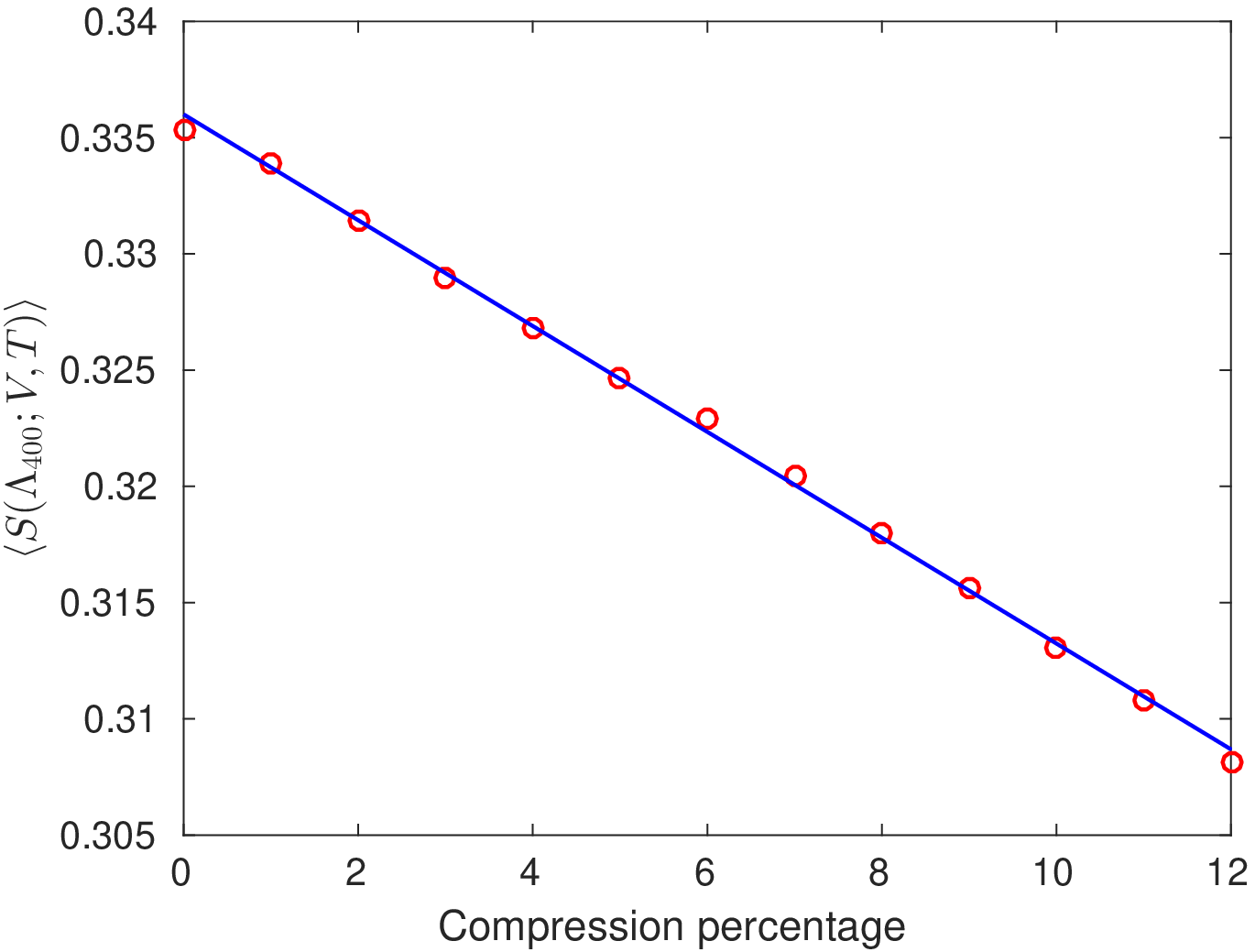}} \quad
\subfloat[][\emph{}]
{\includegraphics[width=.46 \textwidth]{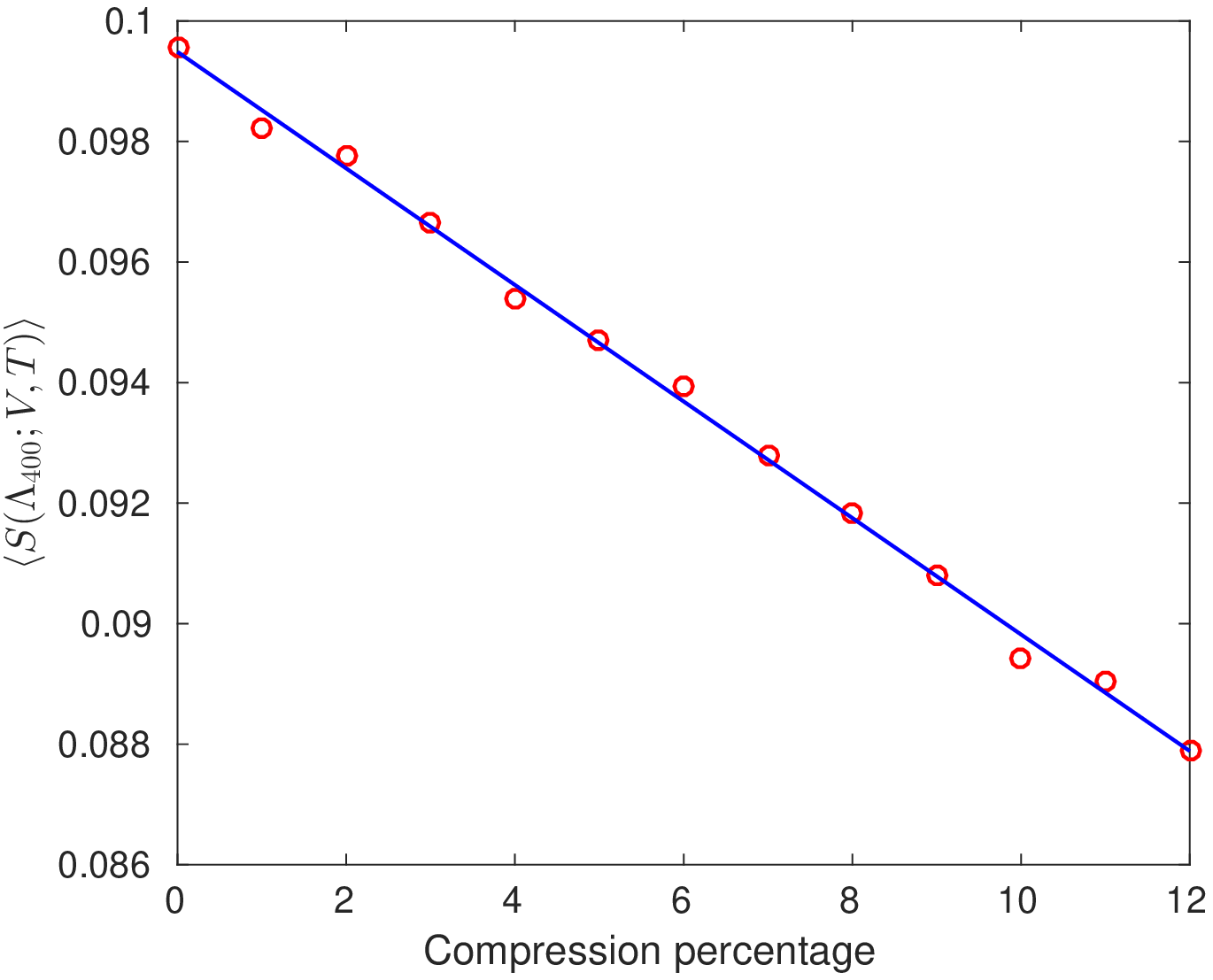}} \quad
{\includegraphics[width=.46  \textwidth]{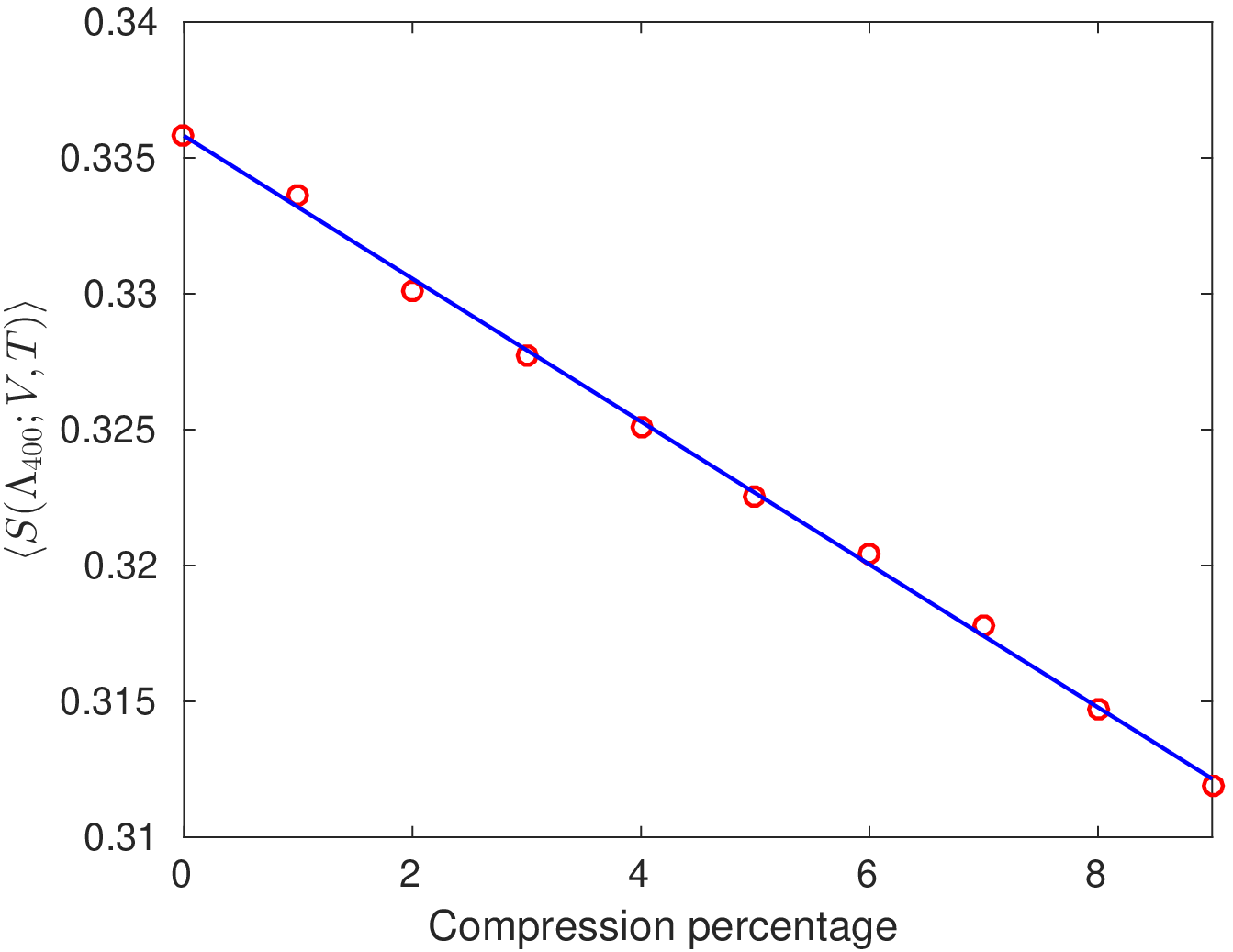}} \quad
\subfloat[][\emph{}]
{\includegraphics[width=.46 \textwidth]{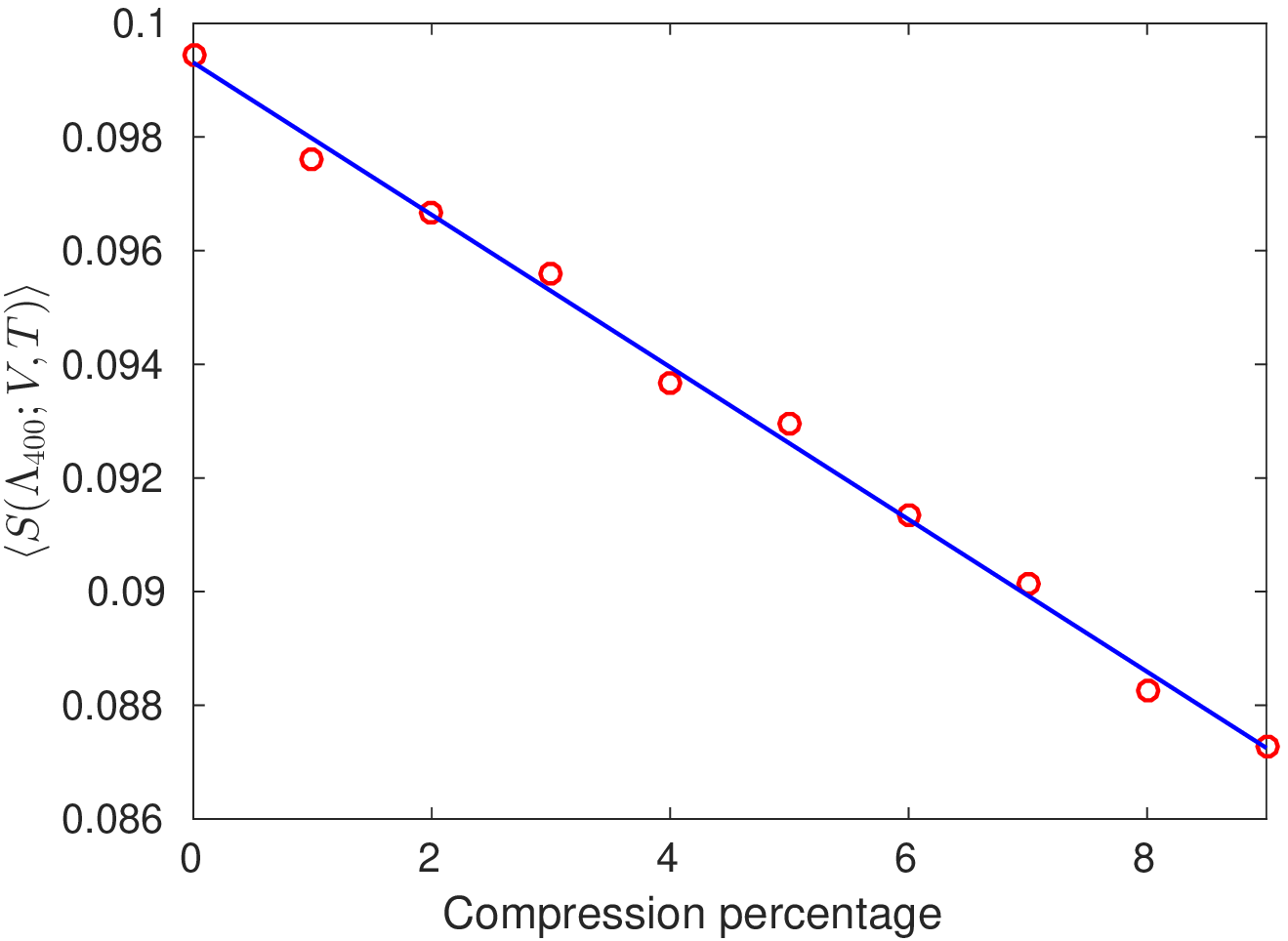}} \quad
\caption{Decay of $\langle S(\Lambda_{400};V,T) \rangle_\Omega$ for different compression percentages and 
different $r$ compared with a linear regression y=ax+b. a) $r$=0.6 and $\lambda\in[0.4,0.6]$, regression coefficient $a=-2.276\cdot 10^{-3}$  St error $2.595\cdot 10^{-5}$, b=0.336  St error 1.835$\cdot 10^{-4}$. b) $r$=0.6 and $\lambda\in[0.1,0.9]$, regression coefficient $a=-9.667\cdot 10^{-4}$ St error $1.508\cdot 10^{-5}$, $b=9.949\cdot 10^{-2}$ St error$ 1.066\cdot 10^{-4}$ c)$r$=1 and $\lambda\in[0.4,0.6]$, regression coefficient 
$a=-2.629\cdot 10^{-3}$ St error 3.48$\cdot 10^{-5}$, b=0.336  
St error 1.835$\cdot 10^{-4}$. d)$r$=1 and $\lambda\in[0.1,0.9]$, regression coefficient $a=-1.340\cdot 10^{-3}$ St error $3.047\cdot 10^{-5}$, $b=9.949\cdot 10^{-2}$ St error$ 1.066\cdot 10^{-4}$.}
\label{compressionistima}
\end{figure}

Figure \ref{compressionistima} shows the dependence of $\langle S(\Lambda_{400};V,T)\rangle$ on 
the compression percentage. More precisely, Fig \ref{compressionistima}(a) and \ref{compressionistima}(b) corresponds to Fig \ref{compressionifigure}(a) and Fig \ref{compressionifigure}(b), while Fig \ref{compressionistima}(c) and Fig \ref{compressionistima}(d) refer to the same setting, but r=1.
It is evident that the reduction of the transmission coefficient is linear as 
a function of the compression factor.
Furthermore table \ref{tabella} allows us to conclude that 
 for given disorder, the absolute value of the rate of decrease, $a$,
increases with $r$, the fraction of compression attributed to the insulator. For fixed $r$,  the absolute value of $a$ decreases if the disorder is higher. The increment of $a$ for growing $r$
means that the increment of the potential height is more significant than the 
reduction of $L_{ins}$. 
\begin{table}[h]
\centering
\begin{tabular}{|c|c|r|}
\hline 
$\rho(\lambda)$ & $r=1$ & $r=0.6$\\
\hline
[0.4;0.6] & $-2.629\cdot 10^{-3}$ & $-2.276\cdot 10^{-3}$\\
\hline
[035;0.65] & $-2.589\cdot 10^{-3}$ & $-2.126\cdot 10^{-3}$\\
\hline
[0.3;0.7] & $-2.381\cdot 10^{-3}$ & $-1.812\cdot 10^{-3}$\\
\hline
[0.1;0.9] & $-1.340\cdot 10^{-3}$ & $-9.667\cdot 10^{-4}$\\
\hline
\end{tabular}
\caption{For fixed $r$, the absolute value of $a$ decreases if the disorder is higher, meaning a minor variation of $\langle S(\Lambda_{400};V,T)\rangle$ as the compression percentage grows. Similarly, the absolute value of $a$ decreases if $r$ decreases at a constant disorder degree.}
\label{tabella}
\end{table}

\begin{figure}[h]
\centering
\subfloat[][\emph{}]
{{\includegraphics[width=.485\textwidth]{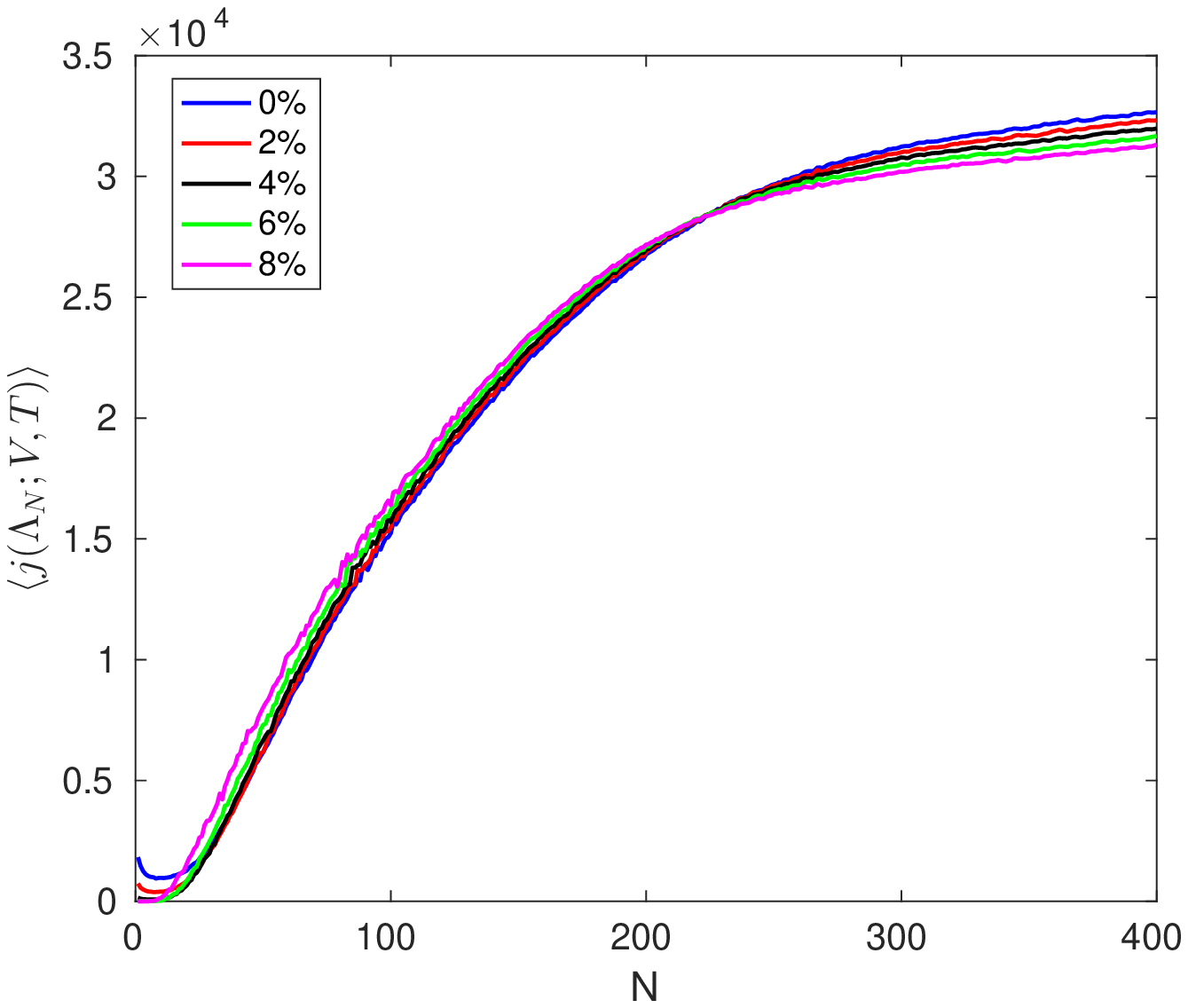}}} \quad
\subfloat[][\emph{}]
{{\includegraphics[width=.485\textwidth]{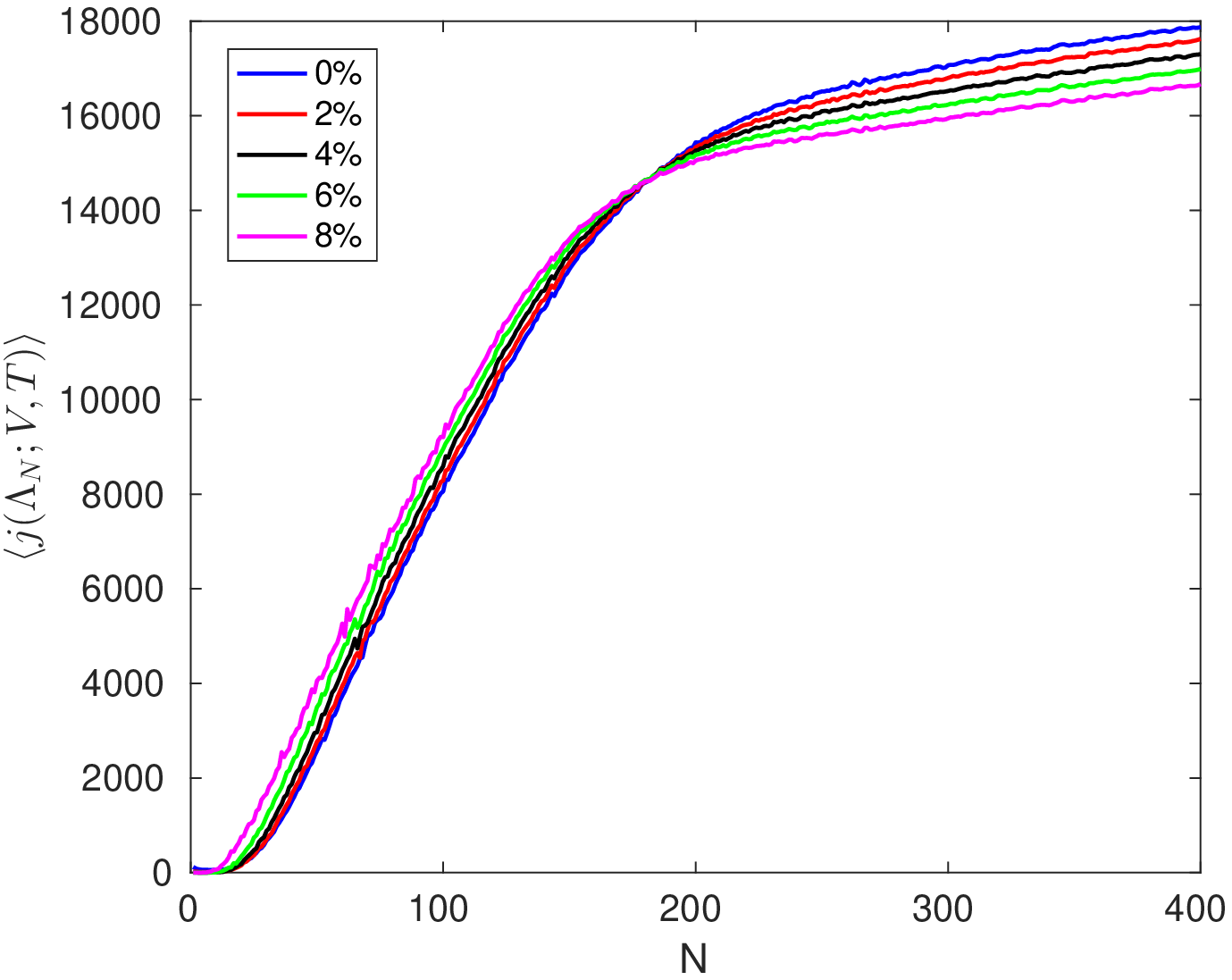}}}
\caption{Behaviour of $\langle j(\Lambda_N;V,T)\rangle$ as $N$ increases for the first model of compression with $r=0.6$ and different compression percentages. Left panel $T=300K$, right panel $T=170K$.}
\label{fig_current}
\end{figure}

Figure. \ref{fig_current} shows the behaviour of the probability current defined by eq. \eqref{current} for different temperatures. Being the problem time independent, the current is constant along the system and it is sufficient to compute it at one of the extreme. For the sake of simplicity, we have fixed $A_0=1$ in the simulations.

We next focus on possible optimal choices for the design of effective sensor devices. From this point of view, it is convenient to examine the relative percentage change of the transmission coefficient under compression, rather than the absolute variation investigated previously. In the following, the relative percentage change is defined as $
\Delta(\beta)=\frac{|\langle S(\Lambda_{N},\beta)\rangle -\langle S(\Lambda_{N},0)\rangle|}{\langle  S(\Lambda_{N},0)\rangle}$, where $\beta\in [0,100]$ is the compression percentage. In particular, we look for good choices for the number of barriers $N$ and for the disorder degree, in order to have a high sensitivity to compression, i.e. large relative percentage change $\Delta(\beta)$ under compression.

Guided by the behaviour of $\langle S(N)\rangle$ described by figures \ref{compressionifigure}, we have considered three possible optimal choices for the variable $N$, that are {\em a)}\  small number of barriers, $N\approx 10$; {\em b)}\ high number of barriers, $N=400$;{\em c)}\ intermediate number of barriers, $N\approx 150$.
 
Table \ref{variazionirelative} (see appendix \ref{sec:appendix2})  summarizes the most interesting values of $\Delta(\beta)$ for these different $N$, disorder degree and ratio $r$. The results show that even tough for $N=400$ we have the maximum absolute drop of the transmission coefficient under compression, the maximum relative drop is attained for smaller numbers of barriers. In fact, especially for strong disorder degree, we have that in the range $N\approx 100\sim 200$, $\Delta(\beta)$ is significantly larger than for $N=400$. For very low $N$, we have a large relative drop of $S$ which might be in theory exploited. Nevertheless since $S$ is very small in absolute value, there might be difficulties to measure the corresponding low currents. Considering the ratio, the higher is $r$, the greater is $\Delta(\beta)$ as we would expect. For $r=1$ and $\beta=10$ results are not shown, because that corresponds to a negative insulator length, cf. \eqref{equazionicompressione}.

We observe that the higher the disorder, the higher the relative percentage change $\Delta(\beta)$, if the other parameters are fixed. Therefore in spite of all the other possible choices, randomness enhances the sensitivity to compression. 

\begin{figure}[h]
\centering
\subfloat[][\emph{}]
{\includegraphics[width=.45 \textwidth]{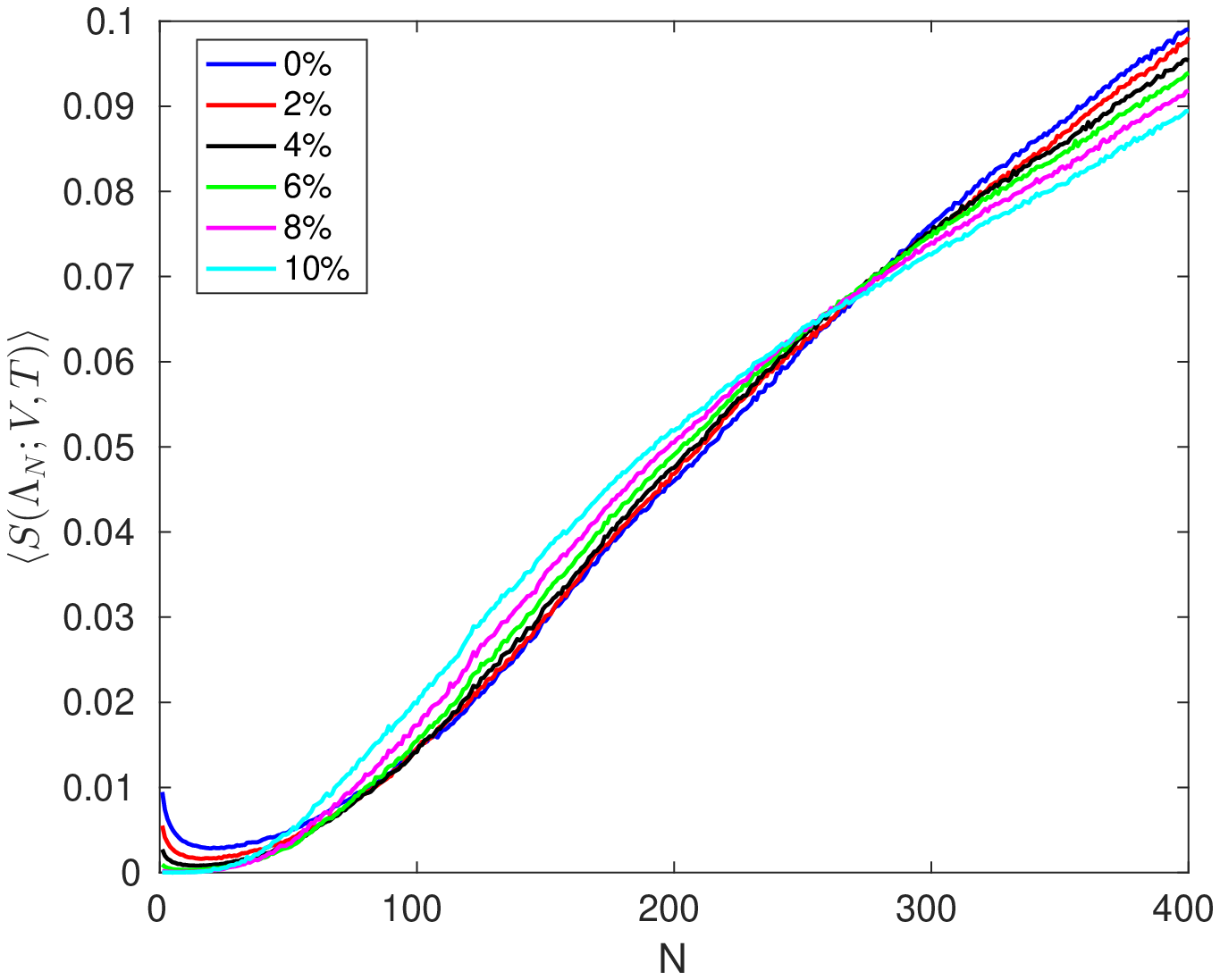}} \quad
\subfloat[][\emph{}]
{\includegraphics[width=.45\textwidth]{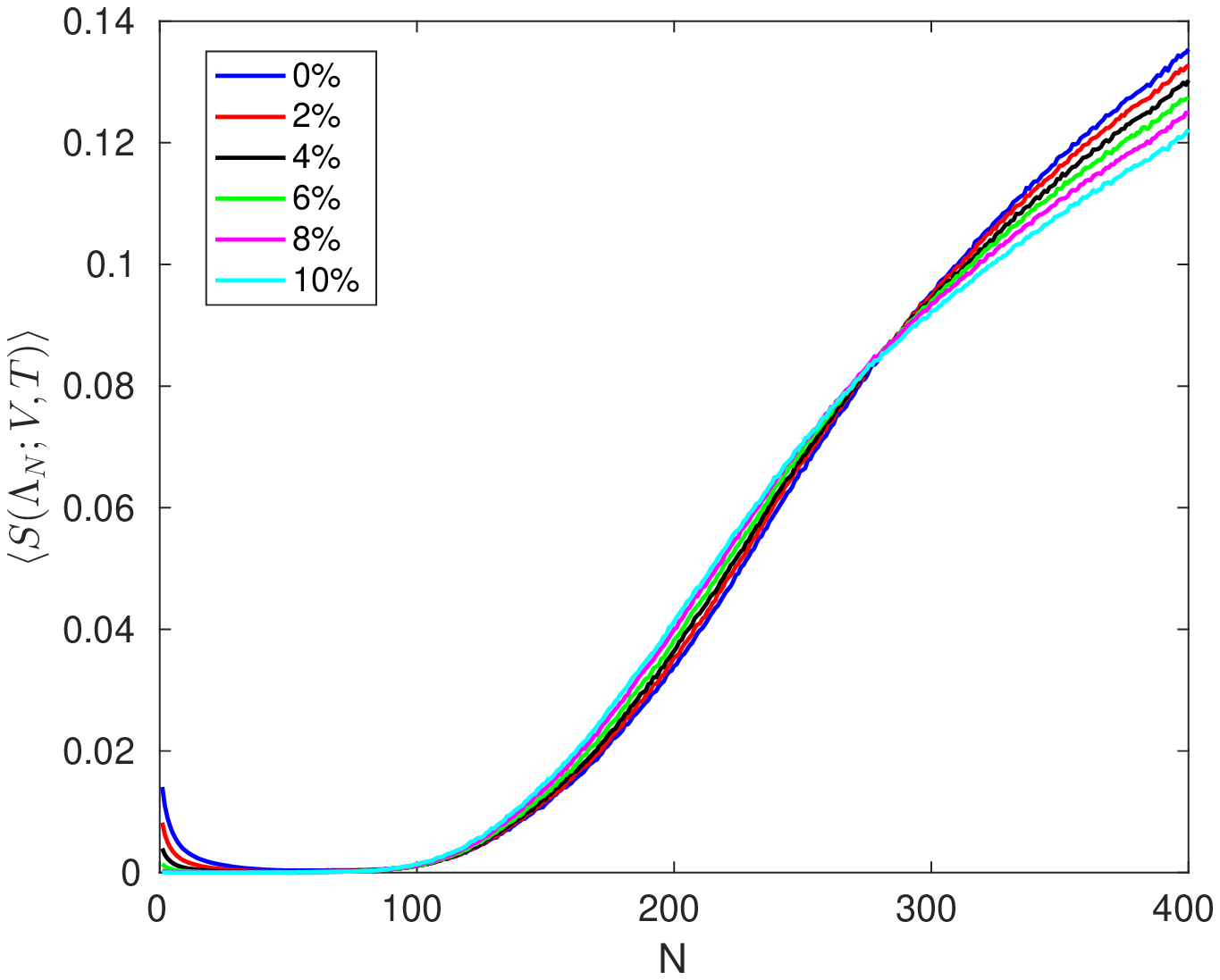}}\quad
\caption{\label{tuttodisordine}Comparison of the transmission coefficient. On the left side disorder is present only in the barrier widths. On the right the barriers widths are strongly disordered, while the wells widths are weakly disordered}
\end{figure}

For this reason, we have also simulated a system in which not only the barriers but also the wells are random. With $\lambda \in [0.1,0.9]$, and a weak disorder for the wells $\delta \in [0.4,0.6]$. Fig \ref{tuttodisordine} shows the behaviour of the transmission coefficient for this system. We observe that the behaviour changes since $\langle S\rangle$  is flat and almost vanishing for $N<100$ and then it grows quickly suggesting a faster rate of convergence to the periodic case than the fixed wells width case.
Nevertheless the relative percentage change remains similar, even tough the crossover zone restricts, as well as the interval of moderate $N$ values for which the relative percentage change is significant. Therefore, introducing randomness in the wells widths does not appear to improve the sensitivity.

All things considered, the optimal design choice for a compression sensor whose barriers height grows linearly with compression, requires a number of barriers $N\approx 100\sim 200$, strong disorder only for the barriers width and ratio $r$ close to one. Nevertheless the absolute variation of the transmission coefficient is small.

\subsection{Generalized compression model}
The numerical results of section \ref{sec:linearcompression} show that a model of compression that preserves the area of the potential barriers produces 
a limited decrease of the transmission coefficient under compression. We therefore propose and numerically
test another possibility. In particular we consider the following rule for the potential height:

\begin{equation}\label{modellocompressione}
V_{compr,a}=V\left(\frac{L(1-\alpha)}{L_{compr}(1-\alpha_{compr})}\right)^2=V\left(\frac{L(1-\alpha)}{L_{ins,compr}}\right)^2=V\left(\frac{L(1-\alpha)}{L_{ins}-L\cdot r\cdot \frac{\beta}{100}}\right)^2
\end{equation}
where $\beta$ is the compression percentage. In this case, the potential increases as  $\frac{1}{(C-\frac{\beta}{100})^2}$ for $\frac{\beta}{100}\rightarrow C$, where $C=\frac{L_{ins}}{L\cdot r}$.

\begin{figure}[h]
\centering
\subfloat[][\emph{}]
{\includegraphics[width=.43 \textwidth]{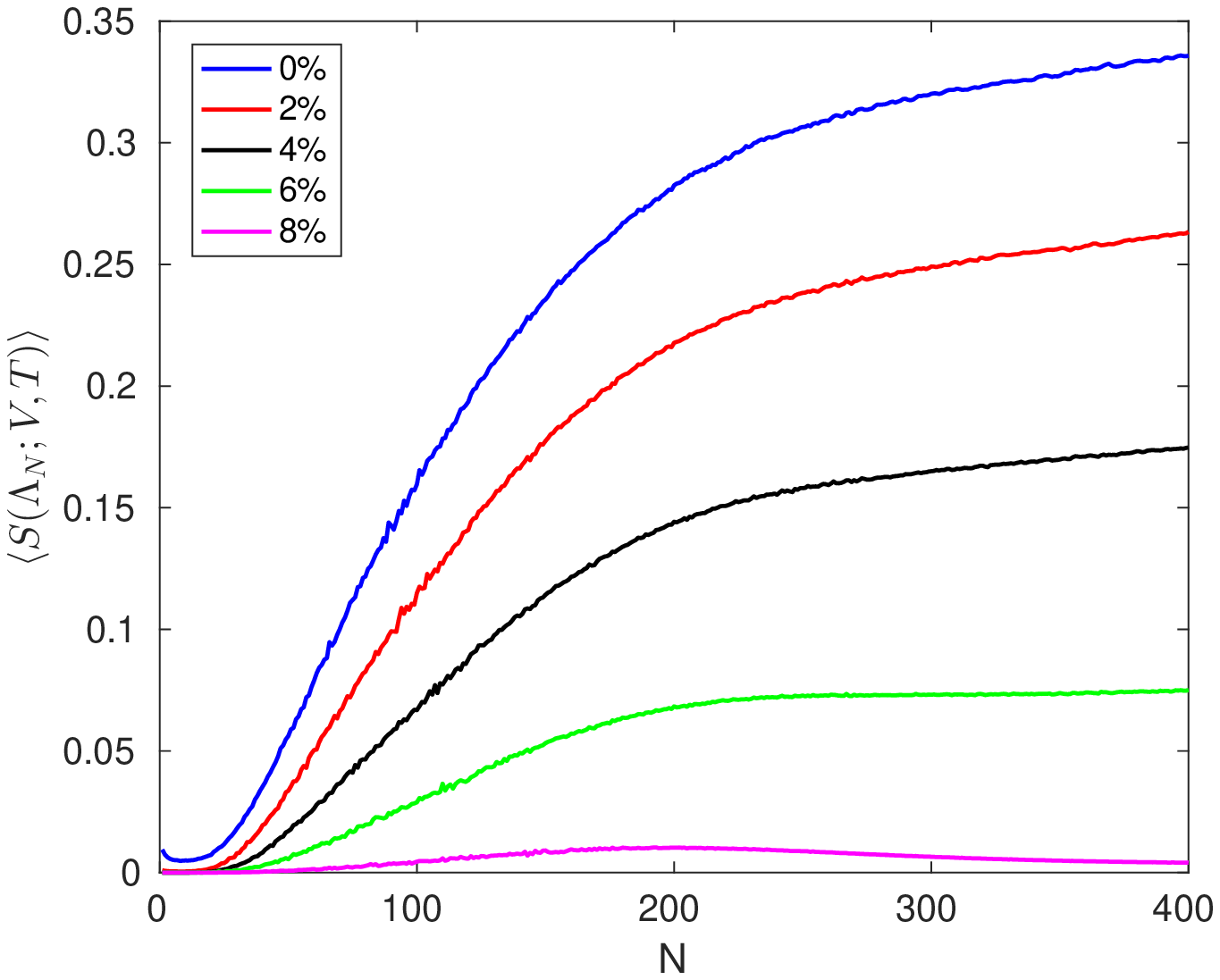}} \quad
\subfloat[][\emph{}]
{\includegraphics[width=.45\textwidth]{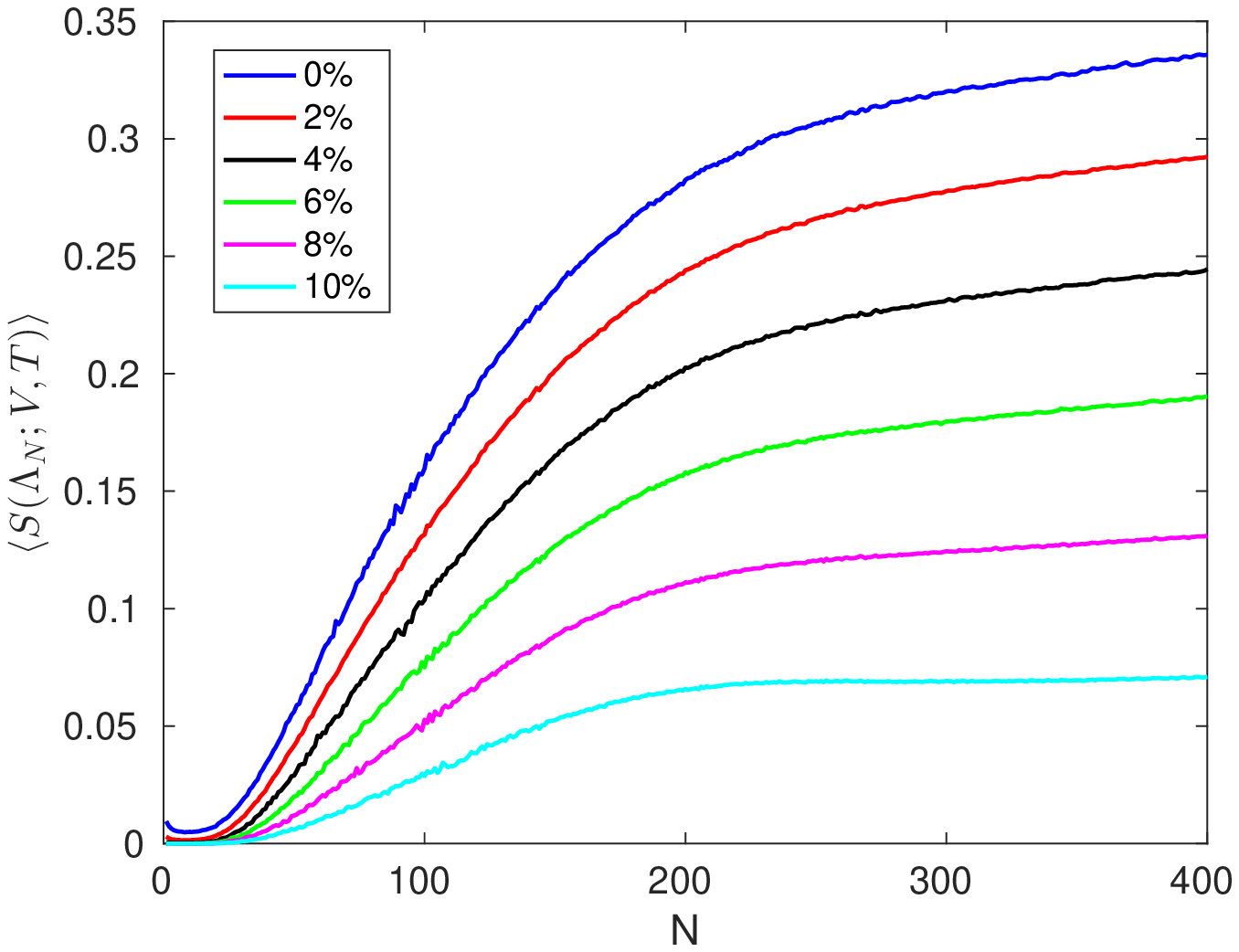}}\quad
\caption{\label{compressionequadrata}Plot of $\langle S(\Lambda_N;V,T)\rangle $ as a function of $N$. 
Panel a) $r$=1 and $\lambda\in[0.4,0.6]$ b) $r$=0.6 and $\lambda\in[0.4,0.6]$ }
\end{figure}

Again, it is to be remarked that for every power $p$, the rule
$$
V_{compr}=V  \left(\frac{L(1-\alpha)}{L_{ins}-L\cdot r\cdot \frac{\beta}{100}}\right)^{p}
$$ 
does not introduce any disorder in the potential heights. 
Taking $p >1$, compression makes the potential increase significantly more than in the case analysed in the previous section, thus we expect the transmission 
coefficient to drop much faster as a function of disorder. This is confirmed by Fig.\ref{compressionequadrata}.

\begin{figure}[ht]
\centering
\subfloat[][\emph{}]
{\includegraphics[width=.45\textwidth]{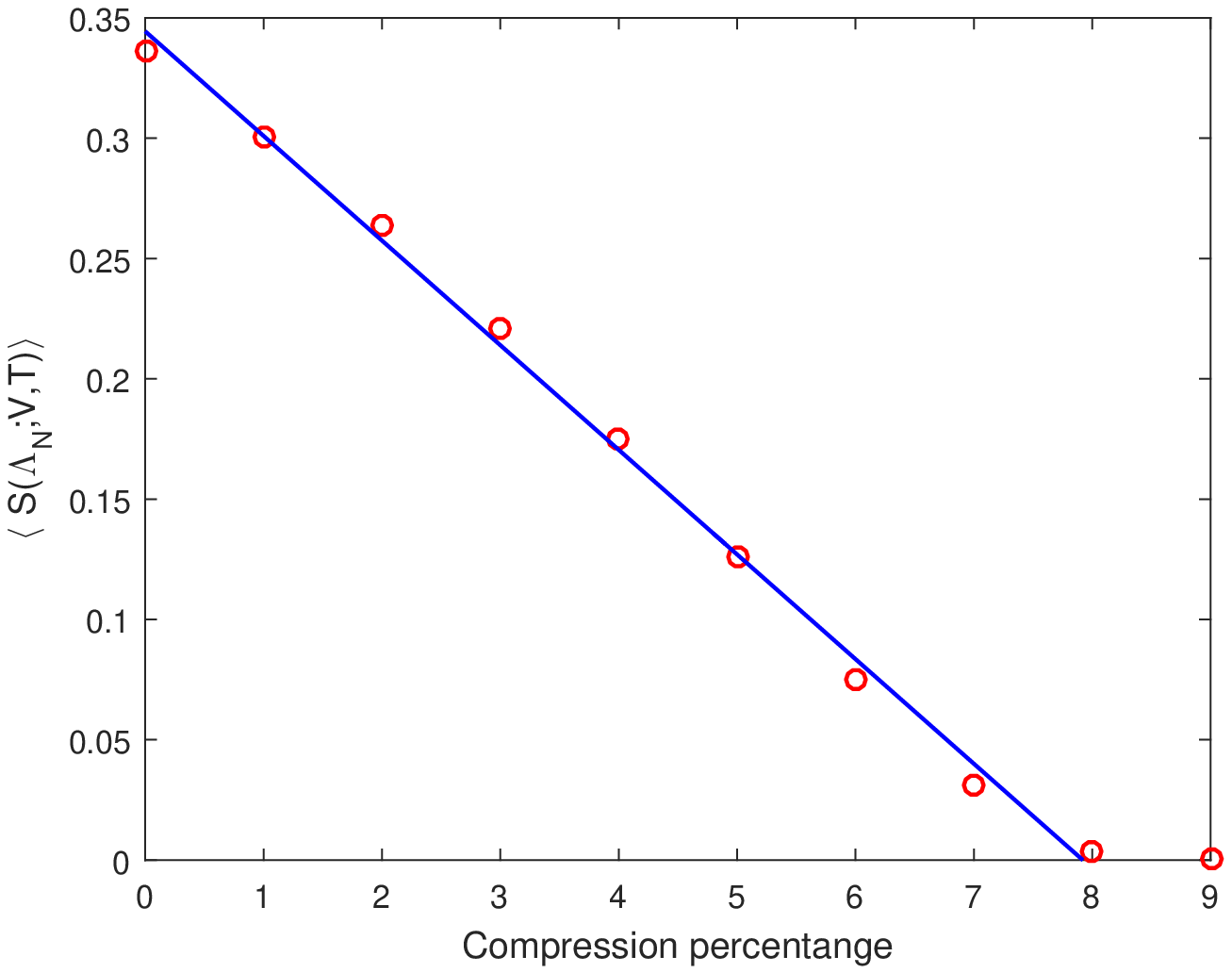}} \quad
\subfloat[][\emph{}]
{\includegraphics[width=.45  \textwidth]{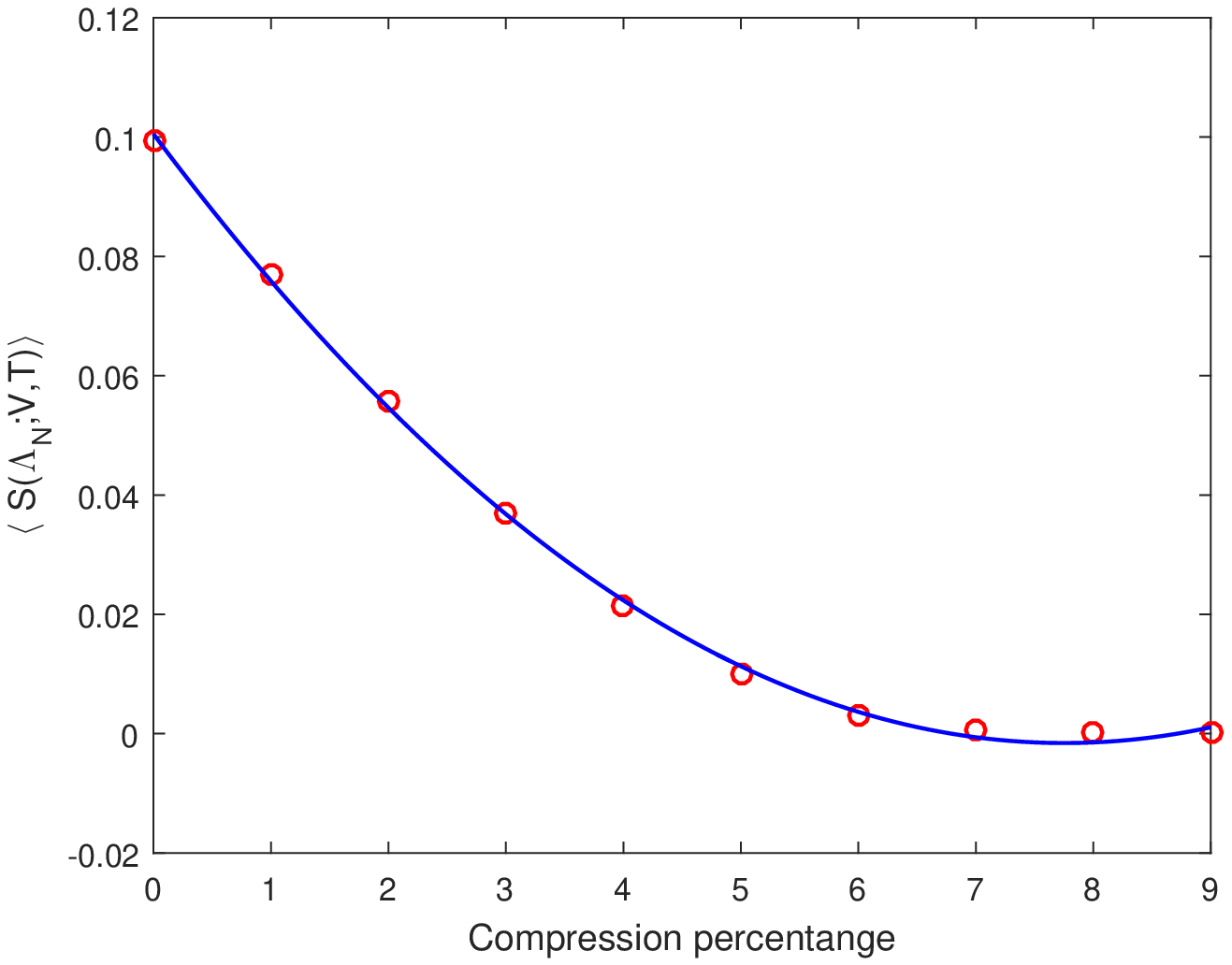}} \quad
\subfloat[][\emph{}]
{\includegraphics[width=.45  \textwidth]{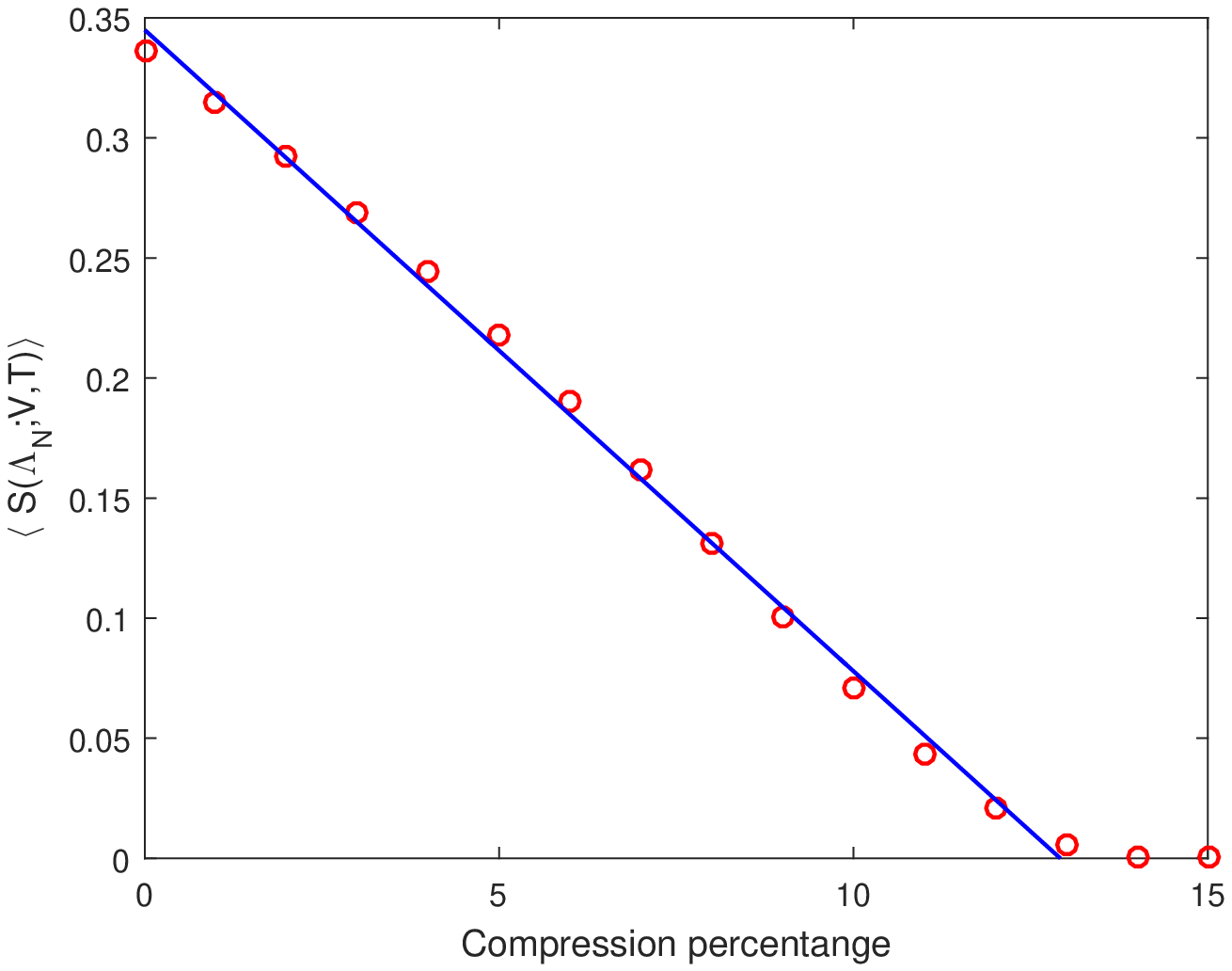}} \quad
\subfloat[][\emph{}]
{\includegraphics[width=.45\textwidth]{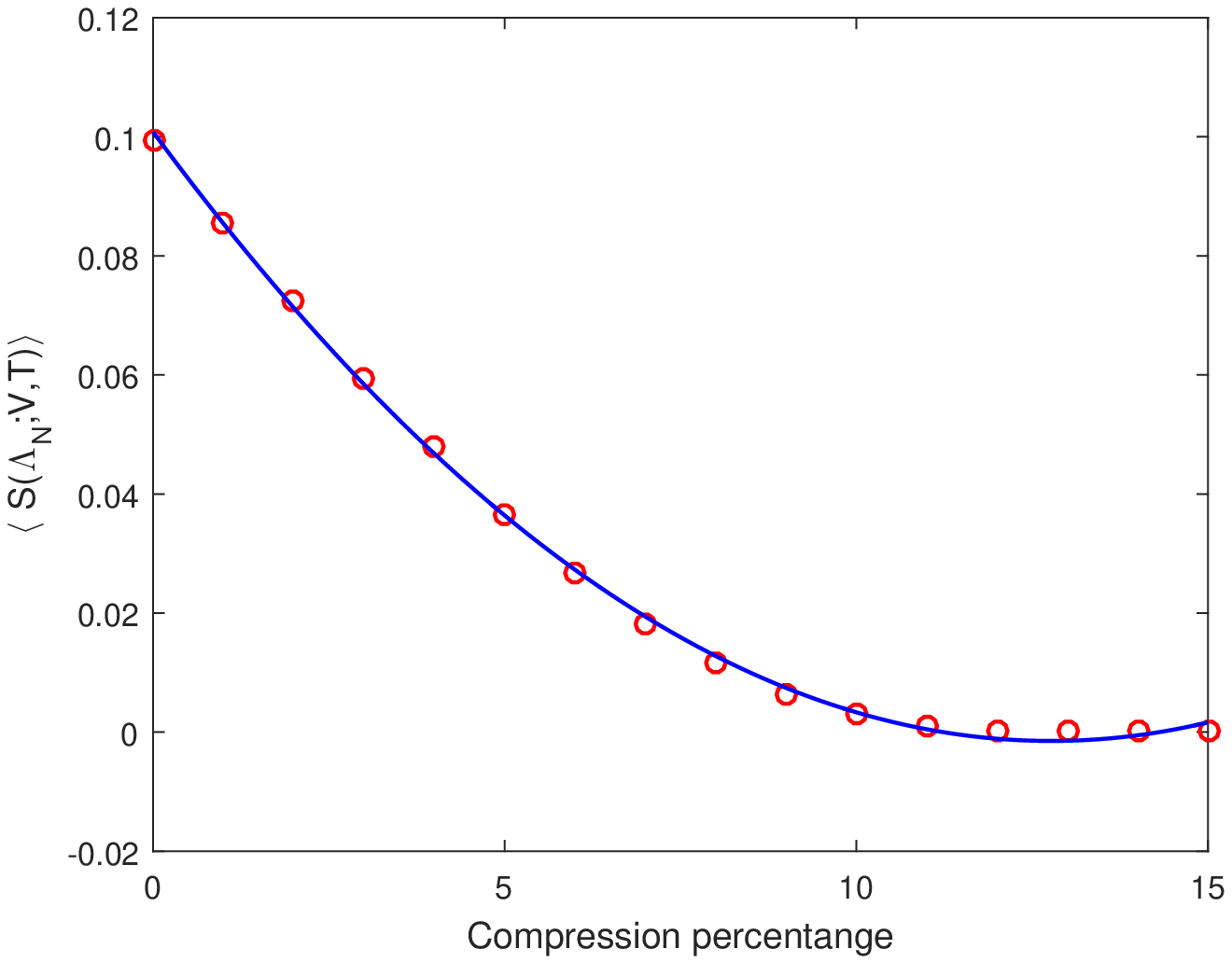}} \quad
\caption{Decay of $\langle S(\Lambda_{400};V,T \rangle$ for different   
compression percentages compared with either linear regression y=ax+b or polynomial regression 
$y=c\cdot x^2 + a\cdot x + b$. a) $r$=1 and $\lambda\in[0.4,0.6]$,$a=-4.34817\cdot 10^{-2}$ St error 9.732$\cdot 10^{-4}$, $b=3.44402\cdot 10^{-1}$  St error 4.6332$\cdot 10^{-3}$ b)r=1 and $\lambda\in[0.1,0.9]$,  $c=+1.698\cdot 10^{-3}$ St error 5.561$\cdot 10^{-5}$,$ a=-2.633\cdot 10^{-2}$  St error 5.199$\cdot 10^{-4}, b=1.005\cdot 10^{-1} $St error $1.005\cdot 10^{-3}. c) $r=0.6 and $\lambda\in[0.4,0.6]$, $a=-2.267\cdot 10^{-2}$  St error $4.04\cdot 10^{-4}$, b=0.344989  St error 3.090$\cdot 10^{-3}$. d) $r$=0.6 and $\lambda\in[0.1,0.9]$, $c=-6.273\cdot 10^{-4}$ St error $1.52\cdot 10^{-5}$, $a=-1.602\cdot 10^{-2}$ St error$ 2.366\cdot 10^{-4}$, b=$1.008\cdot 10^{-1}$ St error=$7.649\cdot 10^{-4}$}
\label{regressionequadrata}
\end{figure}

In Fig.\ref{regressionequadrata} a polynomial regression is shown, to find the decay
rate of the transmission coefficient under compression, with the new potential barriers.
We observe a linear decrease of $S$ with the compression factor for a wide compression range, followed
by a nonlinear, milder decay regime at high compressions. Clearly the absolute value of $a$ is larger than the counterpart for the linear compression model. This indicates that the selection of the material plays a determinant role for the physical properties of the system, and that a non-linear behaviours of the potential height with compression are to be preferred.
\section{Fluctuations and rate functions}
\begin{figure}[h]
\centering
\includegraphics[width=.48\textwidth]{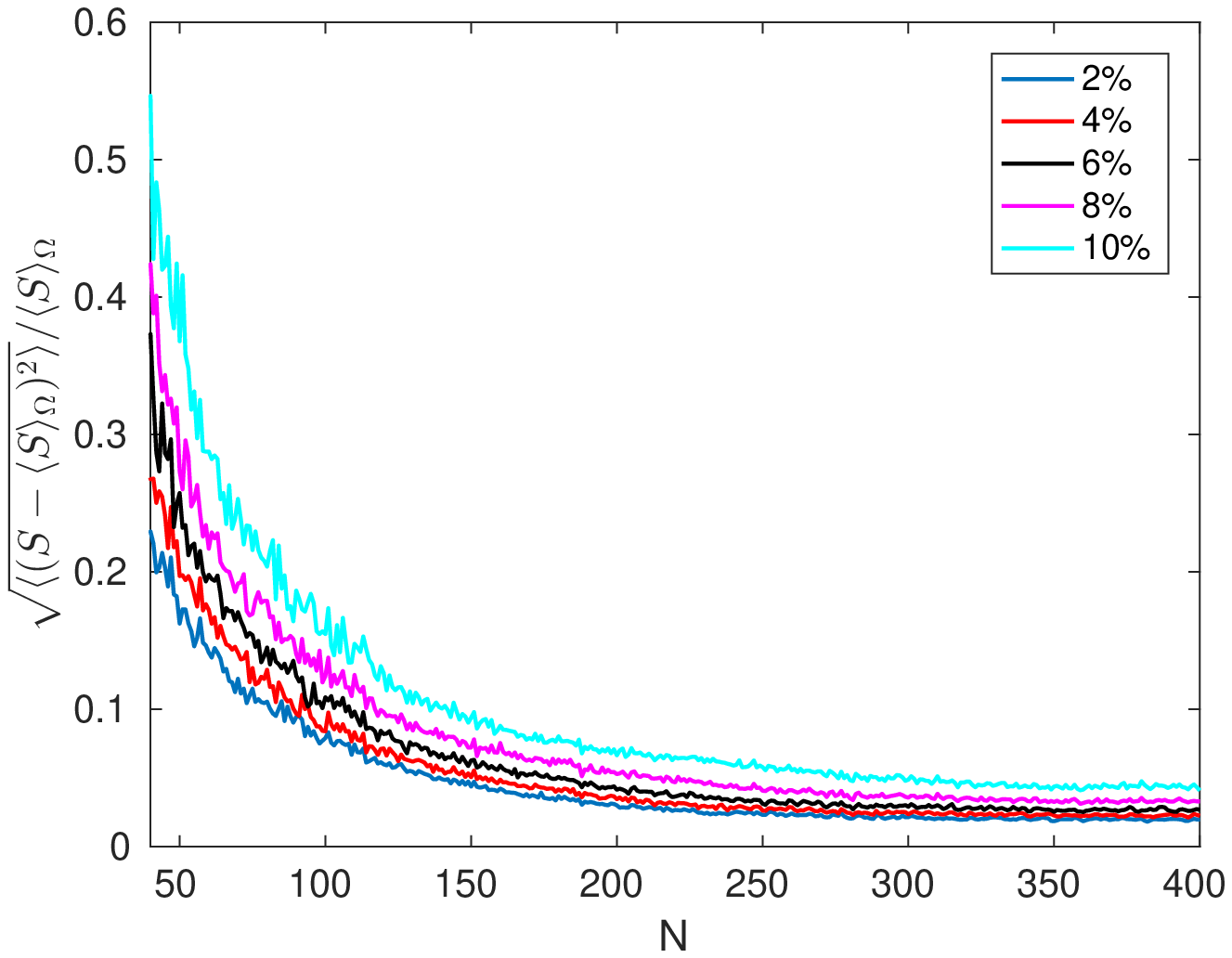} 
\caption{Fluctuations of $S(N,V,T,\Lambda_N)$, r=0.6, quadratic compression model}
\label{fluctuations}
\end{figure}
In \cite{col1}, \cite{col2}, the authors studied the decay of fluctuations as $N$ increases, they identified micro, 
meso and macroscales and checked the validity of a large deviation principle(LDP). It is therefore interesting to 
check how the fluctuations are affected by compression.
In Fig. \ref{fluctuations}, we plot ${\sqrt{\langle (S-\langle S\rangle_{\Omega})^2}\rangle}/{\langle S\rangle_{\Omega}} $ for different percentages of compression. We observe that compression enhances the relative size of  
fluctuations, and it does so more efficiently at small $N$. On the contrary, growing $N$ implies smaller
fluctuations relative size.

Introducing the variable $X_N=\frac{S_N}{\langle S_N \rangle}$, which is the transmission coefficient normalized to 
its expected value, approximated by the empirical mean, and denoting by $\rho_N(X)$ the probability distribution
of $X_N$, we can write
\begin{equation}\label{density}
\langle S \rangle_{\Omega} = \int S \rho_N(X) dX
\end{equation}
\begin{figure}[h]
\centering
\subfloat[][\emph{}]
{\includegraphics[width=.48\textwidth]{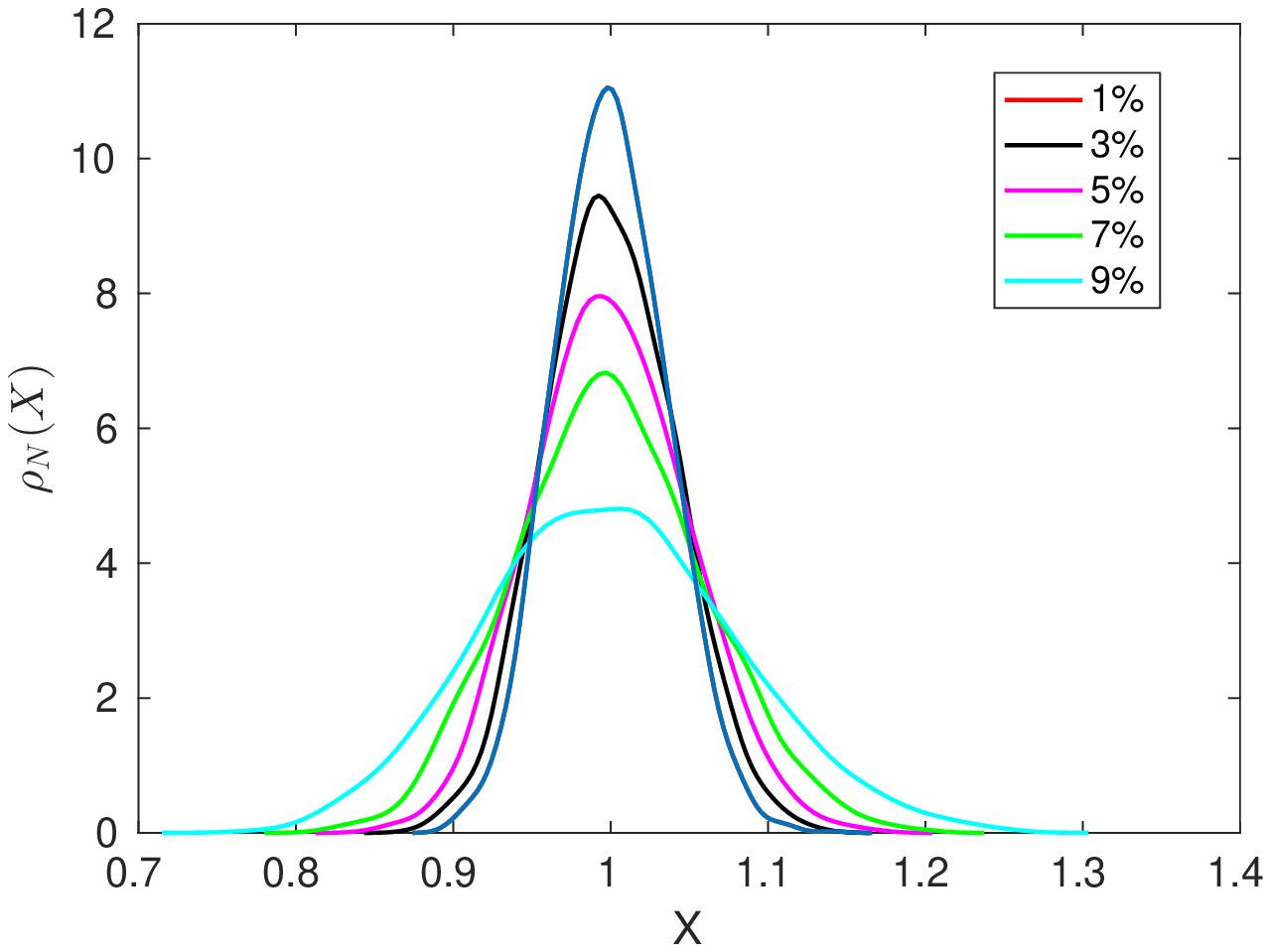}} \quad
\subfloat[][\emph{}]
{\includegraphics[width=.48\textwidth]{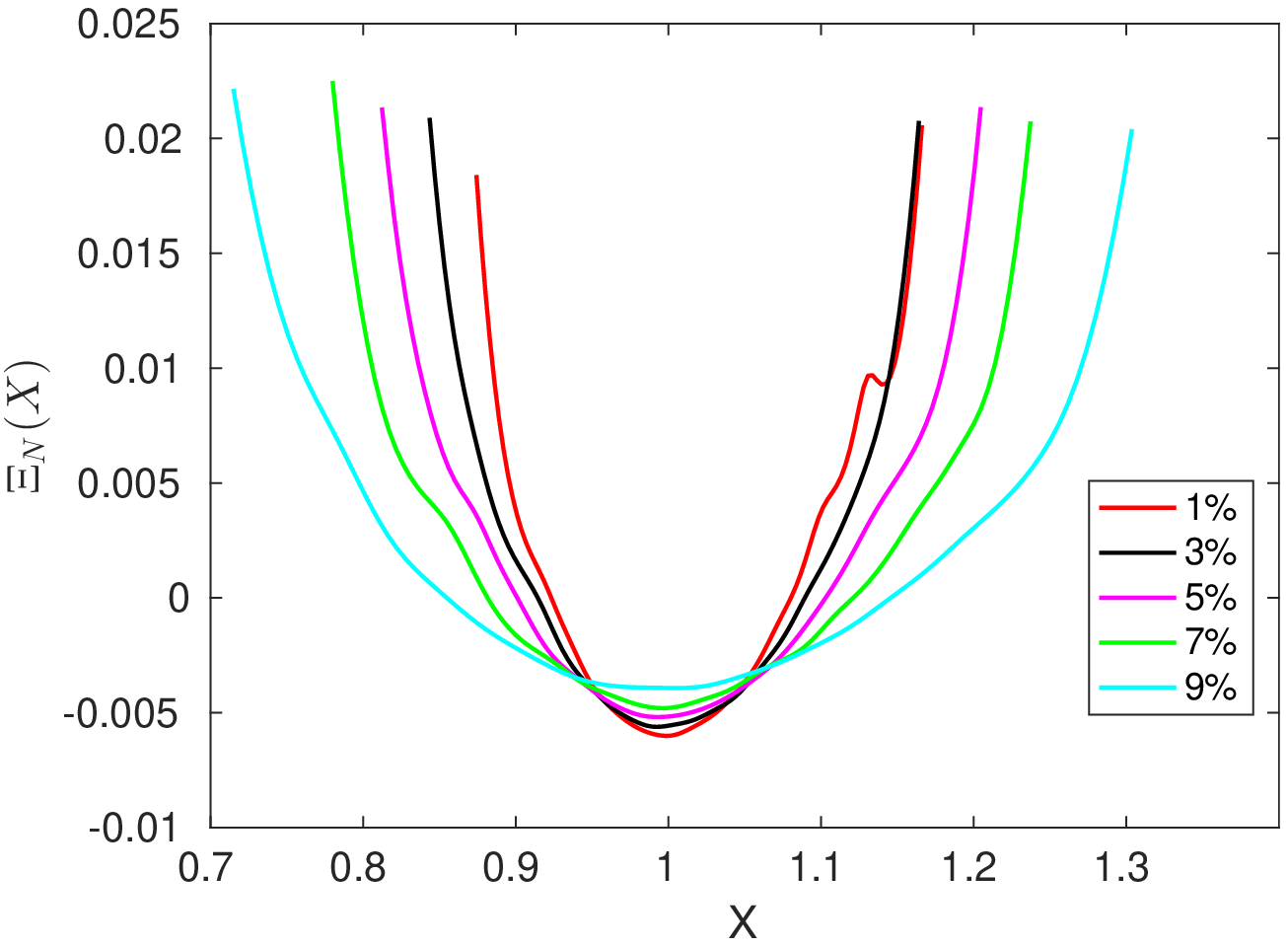}} \quad
\caption{\textbf{a}) Probability density $\rho_N(X)$ in [0.4,0.6] for different compression percentage, r=0.6, quadratic compression model, $N=400$.\textbf{b})Rate functions $\Xi_N(X)$ for same cases of panel (a).}
\label{densityprobability.eps}
\end{figure}
and we may consider now the behaviour under compression of the rate function $\Xi (x)$, \cite{rondoni2003large} \cite{Touchette20091} defined by:
\begin{equation}
\lim_{N\rightarrow \infty} \frac{-\log \rho_N(x)}{N}=\Xi(x)
\end{equation}
Figure \ref{densityprobability.eps}(a) shows that, in accord with Fig.\ref{densityprobability.eps}(b), 
the probability distribution covers a wider range of values 
under larger compression rates.
Therefore, also the properties of the fluctuations can be used to reveal the compression state. Furthermore,
this can be done more efficiently for higher disorder.
\newpage
\section{Conclusion and future developments}
In the present article we have investigated the behaviour of a thermostatted disordered system under compression. 
Our results indicate that for physically relevant $N$, the randomness of the barrier widths leads to a decrease 
of the transmission coefficient, which is more significant for stronger microscopic disorder. 
Considering the large $N$ limit, we have shown numerically that our model behaves similarly to 
the large $N$ limit of the Kronig-Penny model recently studied in \cite{col3} and we have expanded that work considering energy averages. 
We have then shifted our attention to compressed systems, providing a mathematical framework suitable for real cases and amenable to experimental tests. For two compression models we find that compression causes a decrease of $S$. Modifying the degree of freedom $p$, which represents the power law followed by the potential heights under compression,  our numerical simulations show that for quite 
a large interval of compression percentages, the decrease can be assumed to be linear. 
Furthermore for the linear compression model we have extensively investigated the relative percentage change of the transmission coefficient, identifying the best possible configuration for effective sensors.
Eventually we have noticed that compression increases the fluctuations of $S$, as shown by a probability density and rate function estimation. This effect may be used to reveal the compression state of the sample.

\section*{Acknowledgments}
The authors are grateful to M. Colangeli for very useful remarks.
Computational resources were provided by HPC@POLITO (http://hpc.polito.it)

\section*{References}
\bibliography{Main_file_incl_figures}

\begin{thebibliography}{10}
\expandafter\ifx\csname url\endcsname\relax
  \def\url#1{\texttt{#1}}\fi
\expandafter\ifx\csname urlprefix\endcsname\relax\def\urlprefix{URL }\fi
\expandafter\ifx\csname href\endcsname\relax
  \def\href#1#2{#2} \def\path#1{#1}\fi

\bibitem{de2013non}
S.~R. De~Groot, P.~Mazur, Non-equilibrium thermodynamics, Courier Corporation,
  2013.

\bibitem{col1}
M.~Colangeli, L.~Rondoni, Fluctuations in quantum one-dimensional thermostatted
  systems with off-diagonal disorder, Journal of Statistical Mechanics: Theory
  and Experiment 2013~(02) (2013) P02009.

\bibitem{col2}
M.~Colangeli, M.~Pizzi, L.~Rondoni, Current in a quantum driven thermostatted
  system with off-diagonal disorder, Physica A: Statistical Mechanics and its
  Applications 392~(14) (2013) 2977--2987.

\bibitem{anderson}
P.~W. Anderson, Absence of diffusion in certain random lattices, Physical
  review 109~(5) (1958) 1492.

\bibitem{markos2008wave}
P.~Markos, C.~Soukoulis,
  \href{https://books.google.ch/books?id=RSVbFGGfEnsC}{Wave Propagation: From
  Electrons to Photonic Crystals and Left-Handed Materials}, Princeton
  University Press, 2008.
\newline\urlprefix\url{https://books.google.ch/books?id=RSVbFGGfEnsC}

\bibitem{Celardo1}
G.~Celardo, A.~Biella, L.~Kaplan, F.~Borgonovi,
  \href{http://dx.doi.org/10.1002/prop.201200082}{Interplay of superradiance
  and disorder in the anderson model}, Fortschritte der Physik 61~(2-3) (2013)
  250--260.
\newblock \href {http://dx.doi.org/10.1002/prop.201200082}
  {\path{doi:10.1002/prop.201200082}}.
\newline\urlprefix\url{http://dx.doi.org/10.1002/prop.201200082}

\bibitem{Celardo2}
G.~Celardo, A.~Smith, S.~Sorathia, V.~Zelevinsky, R.~Sen’Kov, L.~Kaplan,
  Transport through nanostructures with asymmetric coupling to the leads,
  Physical Review B 82~(16) (2010) 165437.

\bibitem{Celardo3}
G.~Celardo, L.~Kaplan, Superradiance transition in one-dimensional
  nanostructures: An effective non-hermitian hamiltonian formalism, Physical
  Review B 79~(15) (2009) 155108.

\bibitem{soukoulis1981off}
C.~Soukoulis, E.~Economou, Off-diagonal disorder in one-dimensional systems,
  Physical Review B 24~(10) (1981) 5698.

\bibitem{col3}
M.~Colangeli, S.~Ndreca, A.~Procacci, A continuum limit for the kronig--penney
  model, Journal of Statistical Mechanics: Theory and Experiment 2015~(6)
  (2015) P06006.

\bibitem{schwabl}
F.~Schwabl, \href{https://books.google.ch/books?id=pTHb4NK2eZcC}{Quantum
  Mechanics}, Springer Berlin Heidelberg, 2007.
\newline\urlprefix\url{https://books.google.ch/books?id=pTHb4NK2eZcC}

\bibitem{perez}
R.~P{\'e}rez-{\'A}lvarez, R.~Pernas-Salom{\'o}n, V.~Velasco, Relations between
  transfer matrices and numerical stability analysis to avoid the $\omega$d
  problem, SIAM Journal on Applied Mathematics 75~(4) (2015) 1403--1423.

\bibitem{vulpiani}
A.~Crisanti, G.~Paladin, A.~Vulpiani,
  \href{https://books.google.ch/books?id=LafvAAAAMAAJ}{Products of random
  matrices in statistical physics}, Springer series in solid-state sciences,
  Springer, 1993.
\newline\urlprefix\url{https://books.google.ch/books?id=LafvAAAAMAAJ}

\bibitem{izrailev2012anomalous}
F.~M. Izrailev, A.~A. Krokhin, N.~Makarov, Anomalous localization in
  low-dimensional systems with correlated disorder, Physics Reports 512~(3)
  (2012) 125--254.

\bibitem{rondoni2003large}
L.~Rondoni, G.~P. Morriss, Large fluctuations and axiom-c structures in
  deterministically thermostatted systems, Open Systems \& Information Dynamics
  10~(02) (2003) 105--133.

\bibitem{Touchette20091}
H.~Touchette,
  \href{http://www.sciencedirect.com/science/article/pii/S0370157309001410}{The
  large deviation approach to statistical mechanics}, Physics Reports
  478~(1–3) (2009) 1 -- 69.
\newblock \href
  {http://dx.doi.org/http://doi.org/10.1016/j.physrep.2009.05.002}
  {\path{doi:http://doi.org/10.1016/j.physrep.2009.05.002}}.
\newline\urlprefix\url{http://www.sciencedirect.com/science/article/pii/S0370157309001410}

\end{thebibliography}

\appendix
\section*{Appendix}
\subsection{Numerical scheme: the transfer matrix and scattering matrix techniques}
To overcome the numerical difficulties described in section 2, we have developed a numerical scheme that
relies uniquely upon the scattering matrix. If the transfer matrix relates linearly the wave amplitudes
on the left side with the wave amplitudes on the right side, the scattering matrix
relates linearly the amplitudes of wave exiting the barrier potential with the
amplitudes of the wave entering the barrier potential. Therefore, considering a single
barrier, the following relations hold:
\begin{equation}
\begin{pmatrix}
A_4\\
A_5
 \end{pmatrix}=\textbf{M}_3^{-1}\textbf{M}_2\textbf{M}_1^{-1}\textbf{M}_0\begin{pmatrix}
 A_0\\
 A_1
  \end{pmatrix}=\textbf{T}\begin{pmatrix}
 A_0\\
 A_1
  \end{pmatrix}=\begin{pmatrix}
  T_{11} & T_{12}\\
  T_{21} & T_{22}
     \end{pmatrix}\begin{pmatrix}
      A_0\\
      A_1
       \end{pmatrix}
\end{equation}
\begin{equation}
\begin{pmatrix}
A_4\\
A_1
 \end{pmatrix}=\textbf{S}\begin{pmatrix}
 A_0\\
 A_5
  \end{pmatrix}=\begin{pmatrix}
  S_{11} & S_{12}\\
  S_{21} & S_{22}
     \end{pmatrix}\begin{pmatrix}
      A_0\\
      A_5
       \end{pmatrix}
\end{equation}  
It is straightforward to verify that:
\begin{equation}
\mathbf{S}=\begin{pmatrix}
  S_{11} & S_{12}\\
  S_{21} & S_{22}
     \end{pmatrix}=\begin{pmatrix}
       \frac{T_{11}T_{22}-T_{21}T_{12}}{T_{22}} & \frac{T_{12}}{T_{22}}\\
       -\frac{T_{21}}{T_{22}}& \frac{1}{T_{22}}
          \end{pmatrix}
\end{equation}
Since all the components of $\mathbf{T}$ scale at most as $e^{zd}$, $S_{12}, S_{21}, S_{22}$ are bounded. $S_{11}$ might instead explode, because the numerator scales as $e^{2zd}$. Nevertheless, introducing 
\begin{eqnarray}
a_1=(1-\frac{z}{ik})(1-\frac{ik}{z}) \hspace{1.5cm} a_2=(1+\frac{z}{ik})(1+\frac{ik}{z})\\
a_3=(1-\frac{z}{ik})(1+\frac{ik}{z}) \hspace{1.5cm} a_2=(1+\frac{z}{ik})(1-\frac{ik}{z})
\end{eqnarray}
one finds that the leading term of the numerator of $S_{1,1}$ is $(a_1a_2-a_3a_4)e^{2zd}$. 
Since $a_1a_2-a_3a_4=0$, we conclude that all the components of $\mathbf{S}$ are bounded.
Suppose now that the scattering matrix $\hat{\mathbf{S}}$ links linearly the wave amplitudes that enter and exit a sequence of $N$ barriers, while $\mathbf{S}$ describes the scattering process trough the $(N+1)th$ that is added to the system. The following relations allow us to construct a unique scattering matrix for the whole system.
\begin{eqnarray}
A_{4N}=\frac{\hat{S}_{11}S_{11}}{1-S_{12}\hat{S}_{21}} A_0 + \left(\frac{\hat{S}_{11}S_{12}\hat{S}_{22}}{1-S_{12}\hat{S}_{21}}+\hat{S}_{12}\right) A_{4N+1}\\
A_1=\left( S_{21}+\frac{S_{22}\hat{S}_{21}S_{11}}{1-S_{12}\hat{S}_{21}}\right) A_0 +\frac{S_{22}\hat{S}_{22}}{1-S_{12}\hat{S}_{21}} A_{4N+1}
\end{eqnarray}
Once we have the total scattering matrix, it is easy to compute the transmission coefficient through
\begin{equation}
S=\frac{|A_{4N}|^2}{|A_0|^2}=|S_{11}|^2
\end{equation}
The scheme illustrated here has the advantage of being numerically stable and not subjected to overflow problems. 
Nevertheless, this advantage comes at the cost of having to deal with non linear relations, which require a 
greater computational effort than the simpler matrix multiplications of Eq \eqref{transfermatrices}. 
Given the present day computer facilities, this is not a serious hinderance.
\newpage
\subsection{Relative percentange  change under compression}\label{sec:appendix2}
\begin{table}[h]
\centering
\begin{tabular}{|l|l|l|l|l|l|r|}
\hline\hline
Periodic case $N=400,r=0.6$ &   $\beta=2$   &  $\beta=4$ & $\beta=6$ & $\beta=8$ & $\beta=10$  \\
\hline
$\Delta(\beta)$ &   1.08   &  1.95 & 3.29 & 4.65 & 5.65\\
\hline\hline
$\lambda\in [0.4;0.6]$ $N=400,r=0.6$ &   $\beta=2$   &  $\beta=4$ & $\beta=6$ & $\beta=8$ & $\beta=10$  \\
\hline
$\Delta(\beta)$ &   1.16   &  2.54 & 3.72 & 5.17 & 6.64\\
\hline\hline
$\lambda\in [0.4;0.6]$ $N=400,r=1$ &   $\beta=2$   &  $\beta=4$ & $\beta=6$ & $\beta=8$ & $\beta=10$  \\
\hline
$\Delta(\beta)$ &   1.70   &  3.19 & 4.6 & 6.28 & -- \\
\hline\hline
$\lambda\in [0.3;0.7]$ $N=400,r=0.6$ &   $\beta=2$   &  $\beta=4$ & $\beta=6$ & $\beta=8$ & $\beta=10$  \\
\hline
$\Delta(\beta)$ &   1.73   &  3.2 & 4.88 & 6.56 & 8.14\\
\hline\hline
$\lambda\in [0.3;0.7]$ $N=150,r=0.6$ &   $\beta=2$   &  $\beta=4$ & $\beta=6$ & $\beta=8$ & $\beta=10$  \\
\hline
$\Delta(\beta)$ &   1.10   &  2.54 & 4.66 & 6.83 & 9.10\\
\hline\hline
$\lambda\in [0.1;0.9]$ $N=400,r=0.6$ &   $\beta=2$   &  $\beta=4$ & $\beta=6$ & $\beta=8$ & $\beta=10$  \\
\hline
$\Delta(\beta)$ &   1.47   &  4.18 & 5.65 & 7.74 & 10.18\\
\hline\hline
$\lambda\in [0.1;0.9]$ $N=400,r=1$ &   $\beta=2$   &  $\beta=4$ & $\beta=6$ & $\beta=8$ & $\beta=10$  \\
\hline
$\Delta(\beta)$ &   2.78   &  5.81 & 8.14 & 11.24 & --\\
\hline\hline
$\lambda\in [0.1;0.9]$ $N=150,r=0.6$ &   $\beta=2$   &  $\beta=4$ & $\beta=6$ & $\beta=8$ & $\beta=10$  \\
\hline
$\Delta(\beta)$ &   1.98   &  6.00 & 10.71 & 19.10 & 28.17\\
\hline\hline
$\lambda\in [0.1;0.9]$ $N=150,r=1$ &   $\beta=2$   &  $\beta=4$ & $\beta=6$ & $\beta=8$ & $\beta=10$  \\
\hline
$\Delta(\beta)$ &   3.7   &  12.22 & 29.18 & 52.99 & --\\
\hline\hline
$\lambda\in [0.1;0.9]$ $N=100,r=1$ &   $\beta=2$   &  $\beta=4$ & $\beta=6$ & $\beta=8$ & $\beta=10$  \\
\hline
$\Delta(\beta)$ &   5   &  7.92 & 40.76 & 94.88 & --\\
\hline\hline
$\lambda\in [0.1;0.9]$ $N=200,r=1$ &   $\beta=2$   &  $\beta=4$ & $\beta=6$ & $\beta=8$ & $\beta=10$  \\
\hline
$\Delta(\beta)$ &   2.6   &  7.75 & 15.19 & 23.82 & --\\
\hline\hline
$\lambda\in [0.4;0.6]$ $N=10,r=0.6$ &   $\beta=2$   &  $\beta=4$ & $\beta=6$ & $\beta=8$ & $\beta=10$  \\
\hline
$\Delta(\beta)$ &   42.56   &  71.29 & 89.16 & 95.09 & 91.37\\
\hline\hline
\end{tabular}
\caption{Relative percentage variation for different configurations.}
\label{variazionirelative}
\end{table}

\end{document}